\documentclass[aps,prb,twocolumn,showpacs]{revtex4-1}

\usepackage{graphicx}
\usepackage{dcolumn}
\usepackage{bm}
\usepackage{amsmath}
\usepackage{amssymb}
\usepackage{hyperref}
 \usepackage{epstopdf}

\begin{document}
\title{Dissipative Floquet Topological Systems}
\author{Hossein Dehghani$^{1}$}
\author{Takashi Oka$^{2}$}
\author{Aditi Mitra$^{1}$}
\affiliation{$^1$ Department of Physics, New York University, 4 Washington Place, New York, NY 10003, USA\\
$^2$ Department of Applied Physics, University of Tokyo, Hongo 7-3-1, Bunkyo, Tokyo 113-8656, Japan
}
\date{\today}

%%%%%%%%%%%%%%%%%%%%%%%%%%%%%%%%%%%%%%%%%%%%%%%%%%%%%%%%%%%%
%               __   __  ___  __        __  ___            %
%          /\  |__) /__`  |  |__)  /\  /  `  |             %
%         /~~\ |__) .__/  |  |  \ /~~\ \__,  |             %
%                                                          %
%%%%%%%%%%%%%%%%%%%%%%%%%%%%%%%%%%%%%%%%%%%%%%%%%%%%%%%%%%%%
\begin{abstract}
Motivated by recent pump-probe spectroscopies,
we study the effect of phonon dissipation  and potential cooling on the nonequilibrium
distribution function in a Floquet topological state.
To this end, we apply a Floquet kinetic equation
approach to study two dimensional Dirac fermions
irradiated by a circularly polarized laser, a system which is predicted to
be in a laser induced quantum Hall state.
We find that the initial electron distribution shows an anisotropy
with momentum dependent spin textures whose properties are controlled by the
switching-on protocol of the laser. The phonons then smoothen this out
leading to a non-trivial isotropic nonequilibrium distribution which has no memory
of the initial state and initial switch-on protocol, and yet is distinct from
a thermal state. An analytical expression for the
distribution at the Dirac point is obtained that is
relevant for observing quantized transport.
\end{abstract}

\maketitle

\section{Introduction}

Recent years have seen
the emergence of topological states of matter which is
a new way of characterizing materials by the geometric
properties of the underlying band-structure.~\cite{Haldane88,Hasan10,Zhang11,Kane05,Bernevig06} These include
time reversal (TR) breaking integer quantum Hall systems, TR preserving
spin quantum Hall systems or  two-dimensional (2D) topological insulators (TIs),
3D TIs, and their strongly interacting counterparts.~\cite{Senthil14} Another
intriguing class of systems are those
that can show topological behavior only under out of equilibrium
conditions, the main candidate being the Floquet TIs
which arise under periodic driving.~\cite{Oka09,Inoue10,Kitagawa10,Lindner11,Kitagawa11,Lindner13,Gomez13,Podolsky13,Usaj14}

Consider a time periodic Hamiltonian $H(t) = H(t+T)$ where the
periodicity may be due to an external irradiation by a laser.
Then
the time-evolution over one period can be written as $U(t+T,t)=e^{-i H_F T}$ where $H_F$
is the Floquet Hamiltonian: an effective time-independent Hamiltonian
that captures the stroboscopic time-evolution over one period.~\cite{Shirley65,Sambe73}
Floquet TIs have been mainly described by borrowing concepts from
equilibrium where the topological properties are
extracted by analyzing the spectrum of $H_F$, with the topological phase showing non-zero Chern numbers
and edge-states, though the precise correspondence between the usual equilibrium definition of the
Chern number and the number of edge-states does not always work.~\cite{Rudner13,Erhai14,Kundu14}
Experimentally Floquet TIs
have been realized in a photonic system which is effectively in equilibrium because the periodicity in time is
replaced by a periodicity in position.~\cite{Segev13} A Floquet TI has also been realized in a
periodically modulated honeycomb optical lattice of fermionic atoms,~\cite{Esslinger14} where in the
limit of high frequency of modulations, the Floquet Hamiltonian maps onto the Haldane model.~\cite{Haldane88}

However
Floquet TIs are manifestly out of equilibrium and so raise a unique set
of questions that do not arise in systems in equilibrium, one of them
being the issue of the electron distribution function, critical for determining measurable
quantities.
Obviously the
distribution function, at least in ideal closed quantum systems, will depend on
how the periodic driving is switched on~\cite{Lazarides14,Sentef14,Dalibard14} where any switching-on
protocol breaks time-periodicity.
In addition the occupation probability will be very sensitive to any coupling to
an external reservoir.~\cite{Fertig11,Kundu13,Torres14} Often reservoir engineering can even produce topological
properties absent in the closed system,~\cite{Diehl11,Budich14} which in turn requires new measures for
topological order in open and dissipative systems.~\cite{Uhlmann86,Delgado13,Delgado14a}

The main
aim of this paper is to understand the electron distribution function of Floquet topological systems
by accounting for the initial switching-on protocol of the periodic drive
and accounting for coupling to a reservoir of phonons. We will derive and solve a kinetic equation for the electron
distribution function, and
show that the combined effect of drive and dissipation can stabilize non-trivial steady-states. We will
discuss the signature of these states on spin and angle resolved photoemission (ARPES).

The outline of the paper is as follows, in Section~\ref{model} the model is introduced, in Section~\ref{quench}
physical quantities are calculated for the closed system and for a quench switching-on protocol of the laser.
In Section~\ref{phonons} we generalize to the open system where the electrons are coupled to a phonon reservoir. The rate or kinetic equation
accounting for inelastic electron-phonon scattering in the presence of a periodic drive is derived, the results for
physical quantities at steady-state are
obtained and compared with results for the closed
system. Finally in section~\ref{conclu} we present our conclusions. Derivation of general expressions
for the Green's functions needed for ARPES is given in Appendix~\ref{gf}. Analytic
results can be obtained in the vicinity of the Dirac point for both the closed and the open system, and these
are derived in Appendices~\ref{k0} and~\ref{k0ph} respectively.

\section{Model} \label{model}
We study 2D Dirac fermions
coupled to an external circularly polarized laser, and also coupled to a bath of phonons. The Hamiltonian is,
\begin{eqnarray}
H = H_{\rm el} + H_{\rm ph} + H_c
\end{eqnarray}
where (setting $\hbar=1$)
\begin{eqnarray}
&&H_{\rm el}= \sum_{\vec{k}=\left[k_x,k_y\right],\sigma,\sigma'=\uparrow,\downarrow}
c_{\vec{k}\sigma}^{\dagger}\left[\vec{k} + \vec{A}(t)\right]\cdot\vec{\sigma}_{\sigma \sigma'}c_{\vec{k}\sigma'}
\end{eqnarray}
$c^{\dagger}_{\vec{k}\sigma},c_{\vec{k}\sigma}$ are creation, annihilation operators for the Dirac fermions whose velocity $v=1$,
$\vec{\sigma}= \left[\sigma_x,\sigma_y\right]$ are the Pauli matrices,
$\vec{A}=\theta(t)A_0\left[\cos(\Omega t),-\sin(\Omega t)\right]$ is the circularly polarized laser which has
been suddenly switched on at time $t=0$, we will refer to this switch-on protocol as a quench. This model plays a central role
in the study of Floquet topological states where the
circularly polarized laser generates a mass term $m\sigma_z$
in the Floquet Hamiltonian $H_F$,\cite{Oka09,Kitagawa11} with $m=\frac{A_0^2}{\Omega}$
in the high-frequency limit of $A_0/\Omega \ll 1$. This implies
a Hall conductivity $\sigma_{xy}={\rm sign}(m) e^2/2h$ provided a zero
temperature equilibrium distribution at half-filling is realized.
$H_F$ is also the continuum limit of the Haldane model~\cite{Haldane88}
which is an example of a TI (Chern insulator), and was recently realized experimentally
using optical lattices.~\cite{Esslinger14}
Gapless surface states of a 3D TI are also modeled by
the Dirac Hamiltonian, and were recently studied
by pump-probe spectroscopy,~\cite{Gedik13,Onishi14} while
laser induced Hall effect and chiral edge states
are being experimentally studied in graphene.~\cite{Karch10,Karch11}
\begin{figure}
\centering
\includegraphics[height=8cm,width=8cm,keepaspectratio]{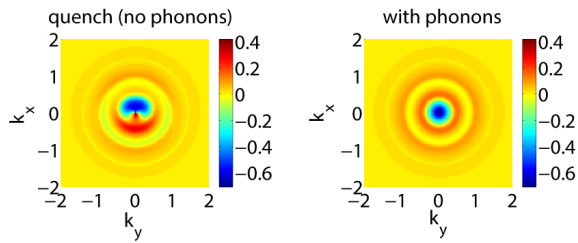}
\caption{(color online) Contour plots for the time-averaged spin density $P_z(k_x,k_y)$.
Left panel: Without phonons and
for a quench. Right panel:
At steady-state with phonons. $A_0/\Omega=0.5$, $\lambda^2D_{\rm ph}=0.1\Omega, T=0.01\Omega,\Omega=1$.
}
\label{fig1}
\end{figure}

Dissipation affects the electron distribution
and thus the topological signatures.
Here we consider dissipation due to coupling to 2D phonons
\begin{eqnarray}
H_{\rm ph}=\sum_{q,i=x,y}\left[\omega_{qi}b_{qi}^{\dagger}b_{qi}\right]
\end{eqnarray}
For now we do not specify whether we have acoustic or optical phonons, and hence the
particular form of the dispersion $\omega_{qi}$. We will specify this when presenting
our results.
The electron-phonon coupling is
\begin{eqnarray}
&&H_c = \sum_{\vec{k},q,\sigma,\sigma'}
c_{\vec{k}\sigma}^{\dagger}\vec{A}_{\rm ph}(q)\cdot\vec{\sigma}_{\sigma \sigma'}c_{\vec{k}\sigma'}\\
&&\vec{A}_{\rm ph}(q) = \left[\lambda_{x,q}\left(b_{x,q}^{\dagger}+b_{x,-q}\right), \lambda_{y,q}\left(b_{y,q}^{\dagger}+b_{y,-q}\right)
\right]
\end{eqnarray}
Above we neglect phonon induced
scattering between electrons with different momenta. This simplifies the kinetic equation for
the electron distribution function considerably, without changing the
physics, and is a microscopic way of accounting for a Caldeira-Leggett~\cite{Caldiera81}
type dissipation.
In Section~\ref{quench}, we will first discuss the physics in the absence of the phonons $H_{c}=0$, but accounting for
the sudden switch-on protocol of the laser, presenting results for the steady-state distribution function
and Green's functions, quantities that are measured in ARPES. In Section~\ref{phonons} we will
address how the results get modified due to coupling to phonons.

\section{ Results for the quench and in the absence of phonons}\label{quench}
Suppose that at $t\leq 0$, there is no external irradiation, and the
electrons are in the ground-state, {\sl i.e.}, all states below the Dirac point are occupied. Thus
the wavefunction right before the switching
on of the laser is
\begin{eqnarray}
&&|\Psi_{\rm in}(t=0^-)\rangle=\prod_{\vec{k}}|\psi_{{\rm in},k}\rangle\nonumber\\
&&|\psi_{{\rm in},k}\rangle= \frac{1}{\sqrt{2}}
\begin{pmatrix}-e^{-i\theta_k}\\ 1 \end{pmatrix}
\end{eqnarray}
where $\theta_k=\arctan\left(k_y/k_x\right)$. The time-evolution after switching
on the laser is
\begin{eqnarray}
|\Psi(t>0)\rangle=U_{\rm el}(t,0)|\Psi_{\rm in}\rangle
\end{eqnarray}
where   $U_{\rm el}(t,t')$ is the time-evolution operator,
\begin{eqnarray}
i\frac{dU_{\rm el}(t,t')}{dt} = H_{\rm el}(t)U_{\rm el}(t,t')\,\,;U_{\rm el}(t,t)=1
\end{eqnarray}
Since we neglect any spatial dependence of the laser field, the system stays translationally invariant. Thus
the dynamics is factorizable between different momenta, $U_{\rm el}(t,t')=\prod_k U_{k}(t,t')$ so that,
$|\Psi(t)\rangle= \prod_{k}|\psi_k(t)\rangle = \prod_k U_k(t,0)|\psi_{{\rm in},k}\rangle$,
where
\begin{eqnarray}
U_k(t,t')= \sum_{\alpha=\pm}|\psi_{k\alpha}(t)\rangle\langle\psi_{k\alpha}(t')|
\end{eqnarray}
$|\psi_{k\alpha}(t)\rangle$ being the exact solution of the Schr\"odinger equation which may be written in terms of
the time-periodic Floquet quasi-modes ($|\phi_{k\alpha}(t+T)\rangle=|\phi_{k\alpha}(t)\rangle$)
and quasi-energies ($\epsilon_{k\alpha}$) as follows,
\begin{eqnarray}
&&|\psi_{k\alpha}(t)\rangle= e^{-i\epsilon_{k\alpha}t}|\phi_{k\alpha}(t)\rangle \nonumber\\
&&\left[H_{\rm el}-i\partial_t\right]|\phi_{k\alpha}\rangle
=\epsilon_{k\alpha}|\phi_{k\alpha}\rangle
\end{eqnarray}
The quasi-energies $\epsilon_{k\alpha}$ represent an infinite ladder of
states where $\epsilon_{k\alpha}$ and $\epsilon_{k\alpha}+m\Omega$, for $m$ any integer,
represent the same physical state corresponding to the
Floquet quasi-modes $|\phi_{k\alpha}(t)\rangle$ and
$e^{im\Omega t}|\phi_{k\alpha}(t)\rangle$. Confusion due to this over-counting can be easily avoided by noting that
in all physical quantities, including the matrix elements for electron-phonon scattering that
enter the kinetic equation, it is always the combination $e^{-i\epsilon_{k\alpha}t}|\phi_{k\alpha}(t)\rangle= |\psi_{k\alpha}\rangle$
that appears, where there are only two distinct states corresponding to $\alpha=\pm$. However while in a typical
two-level system, there is only one energy-scale corresponding to the level splitting, in this problem,
a hierarchy of energy
scales $\epsilon_{k+}-\epsilon_{k-}+ m\Omega$ are possible, although one needs to take care that not all matrix elements
for inelastic processes at these energy-scales may exist. This will be discussed
in more detail below when we present our results.

We can determine the retarded Green's function~\cite{Gedik13b}
\begin{eqnarray}
&&g^R_{\sigma\sigma'}(k,t,t')= -i\theta(t-t') \langle \Psi_{\rm in}|\biggl\{c_{k\sigma}(t),c_{k\sigma'}^{\dagger}(t')\biggr\}|\Psi_{\rm in}\rangle\nonumber\\
&&= -i \theta(t-t')U_{k,\sigma\sigma'}(t,t')\label{grdef}
\end{eqnarray}
and the lesser Green's function,
\begin{eqnarray}
&&g^<_{\sigma\sigma'}(k,t,t')= -i\langle \Psi_{\rm in}|c_{k\sigma}^{\dagger}(t)c_{k\sigma'}(t')|\Psi_{\rm in}\rangle\nonumber\\
&&=-i\sum_{\sigma_1,\sigma_2}\langle\Psi_{\rm in}|c_{k\sigma_2}^{\dagger}c_{k\sigma_1}|\Psi_{\rm in}\rangle
U_{k,\sigma_2\sigma}(0,t)U_{k,\sigma'\sigma_1}(t',0)\nonumber\\\label{glessdef}
\end{eqnarray}
While $g^R$ does not depend on the occupation probability (by not depending on the initial state), $g^{<}$ depends on it.
We perform a Fourier transformation of the Green's functions $g(k,t,t')$
with respect to the time-difference $t-t'$ thus moving to the frequency
$\omega$ space, and all throughout we present results after
time-averaging over the mean time $T_m=(t+t')/2$. Thus in what follows, whenever we denote quantities by the arguments $k,\omega$ alone,
it should be understood that an averaging over mean time $T_m$ has already been performed. General expressions for the
Green's functions are presented in Appendix~\ref{gf} where the averaging procedure over the mean time is also explained. 
\begin{figure}
\centering
\includegraphics[totalheight=5cm]{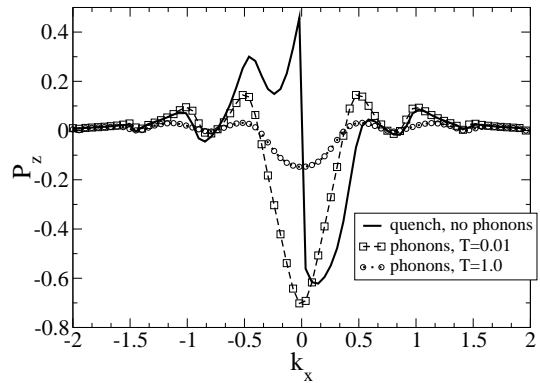}
\caption{Spin density $P_z(k_x,k_y=0)$ for $A_0/\Omega=0.5,\Omega=1.0,\lambda^2D_{\rm ph}$=$0.1\Omega$
for three different cases: for the
quench with no phonons, steady-state with phonons at temperature $T=0.01\Omega$, and $T=\Omega$.
}
\label{fig2}
\end{figure}

We refer to $ig^{<}_{\sigma\sigma}(k,\omega)$
as the spin resolved ARPES spectrum, a key quantity in this work
that can be directly probed in experiments.~\cite{Gedik13}
Note that results for the spectral density
$A={\rm Im}\left[g^R\right]$
have been discussed elsewhere~\cite{Oka09,Gedik13b}, however our results for $g^<$ even in the absence of phonons are new.
We note that number conservation, absence of momentum mixing, and the fact that we are at half-filling
imply the sum rule $\int (d\omega/2\pi)
i\sum_{\sigma}g_{\sigma\sigma}^{<}(k,\omega)= 1$.

\begin{figure}
\centering
\includegraphics[width=9cm,keepaspectratio]{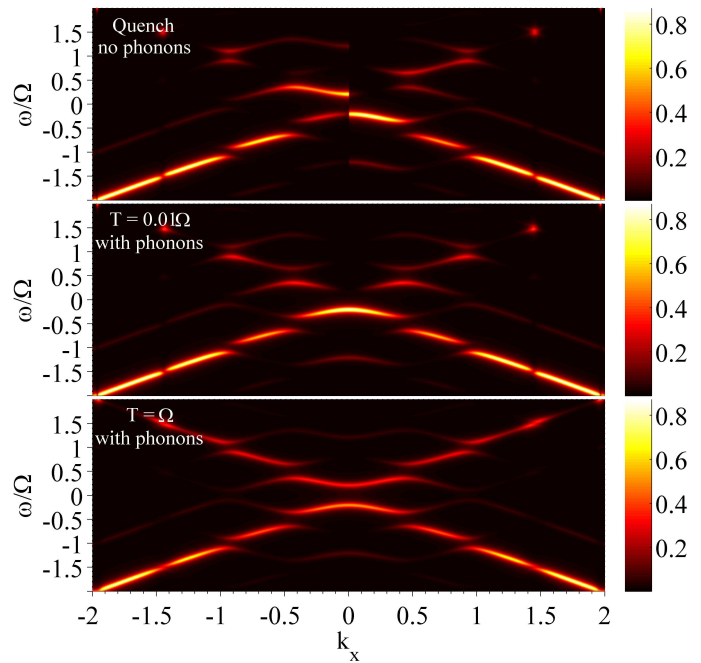}
\caption{(color online) Intensity plot of the spin-averaged ARPES spectrum $iG^{<}_{\rm tot}(k,\omega)/2$ at $k_y=10^{-4}$
for the quench with no phonons (upper panels) and at steady-state
with phonons at temperature $T=0.01\Omega,1\Omega$ (middle and lower panels).
$A_0/\Omega=0.5,\lambda^2D_{\rm ph}=0.1\Omega,\Omega=1.0$.
}
\label{fig3}
\end{figure}

The results for the momentum dependent spin-density $P_{z}(k,T_m)= i \sum_{\sigma}\sigma g^{<}_{\sigma \sigma}(k,T_m,T_m)$
after averaging over $T_m$ is shown
as a contour plot in the left panel of Fig.~\ref{fig1} as well as along the line $k_y=0$ in Fig.~\ref{fig2}.
The circularly polarized laser induces a strongly momentum dependent spin density which
shows oscillations each time the condition
for a photon induced resonance between the Dirac bands $|k| \simeq n \Omega/2$,
where $n$ is an integer, is obeyed. Further, the density is also
anisotropic in momentum space.
The spin averaged ARPES spectrum
$ig^<_{\rm tot}(k,\omega)=i\sum_{\sigma }g_{\sigma\sigma}^{<}(k,\omega)$
is plotted as an intensity plot in Fig.~\ref{fig3}
and its momentum slices in Fig.~\ref{fig4}. Note that the delta-functions are given
an artificial broadening which is such that the heights of the peaks in Fig.~\ref{fig4}
equal the prefactor of the delta function. In other words, $B\pi\delta(\omega -\epsilon_k)$
has been approximated by $\frac{B}{\gamma}\frac{\gamma^2}{\gamma^2 + (\omega-\epsilon_k)^2}$ with
the broadening $\gamma$ 
arbitrary and chosen so that the plots are visible. Moreover
we plot $\gamma g$ (or $\gamma G$ with phonons),
so that the height of the peaks in Fig.~\ref{fig4} equals the prefactor of the
$\delta$-function $B$.

The ARPES spectrum clearly shows the appearance of Floquet bands.
Without phonons, the system is free, and the electron distribution is
given by the overlap $|\langle \phi_{k\alpha=\pm}(0)|\psi_{\rm in, k}\rangle|^2$.
This is a highly non-thermal state that retains memory of
the initial state $|\psi_{\rm in, k}\rangle$, and is not
expected to thermalize.
Just like the spin resolved density, the total
density in Fig.~\ref{fig3}
shows a clear asymmetry under $k_x\rightarrow -k_x$, where this particular anisotropy is determined by the phase of the AC field at 
$t=0^+$. Note that in our case,
initially the gauge field $\vec{A}(t=0^+)=\left[A_0,0\right]$ is entirely along the $\hat{x}$ direction.

The anisotropy can be understood analytically
at $k$=$0$ (see Appendix~\ref{k0} for details),
\begin{eqnarray}
\!\!P_z(k=0,\theta_k)\!\! = \!\!-\frac{2A_0\Omega}{\Delta^2}\cos{\theta_k};\!\Delta= \!\!\sqrt{4 A_0^2 + \Omega^2}
\end{eqnarray}
where $\theta_k$ is the angle along which $k=0$ is approached. This has the same anisotropy as the left panel in Fig.\ref{fig1}.
For the lesser Green's function at $k$=$0$ we obtain,
\begin{eqnarray}
&&ig^{<}_{\sigma\sigma}(k=0,\theta_k,\omega)=2\pi\sum_{\alpha=\pm}\rho_{k=0,\alpha\alpha}^{\rm quench} \nonumber\\
&&\times \biggl[
\left(\frac{\Delta +\alpha \sigma \Omega}{2\Delta}\right)
\delta\left(\omega +\sigma\frac{(\alpha\Delta + \Omega)}{2}\right) \biggr],\nonumber\\
&&\!\!\rho_{k=0,\alpha\alpha}^{\rm quench}= \!\!|\langle \phi_{k=0,\alpha}(0)| \psi_{{\rm in},k=0}\rangle|^2\!\!=
\frac{1}{2}\left(\!1\!-\!\!\frac{2\alpha A_0}{\Delta}\cos{\theta_k}\!\!\right)\nonumber\\\label{glessmain}
\end{eqnarray}
above $\sigma=+/-$ for $\uparrow/\downarrow$.
The analytic expression for $g^<$
shows that for $k=0$, there are exactly four resonances for inelastic scattering,
where the two resonances for spin $\sigma$ occur at $\omega=-\sigma(\Omega\pm\Delta)/2$. Naively one would have expected
infinite number of resonances $\epsilon_{k+}-\epsilon_{k-}+m\Omega$ for integer $m$. The fact that at $k=0$ there are
so few is due to vanishing matrix elements alluded to above. As $k$ increases, more and more resonances appear, however they are
very rapidly suppressed for large $|m|$.

The above location of the resonances also shows that the circularly polarized field
acts as an effective magnetic field along $\hat{z}$,~\cite{Kitagawa11} splitting
the energies of the up and down spin electrons. In particular in the high frequency ($A_0\ll \Omega$) limit,
the lowest energy excitation is $\Delta-\Omega\simeq
2A_0^2/\Omega$ and involves flipping a
spin from $\downarrow$ to $\uparrow$. However, this is still not a typical two level system, as
for $k=0$, there are two energy scales
for energy absorption ($\omega>0$) (and more for $k\neq 0$), rather than just one
energy-scale for energy absorption encountered in a conventional two level system.

The analytic results also show that
the weights are far from thermal, where by thermal we imply
resonances of the form $\delta(\omega-\epsilon_k)n_F(\omega)$, $n_F$ being the Fermi distribution function
at some temperature $T$. Rather the height of the resonances are proportional to amplitude square of the
overlap between the initial wavefunction corresponding to the ground state of the Dirac model, and the wavefunctions 
$|\psi_{k\alpha}\rangle$.
Note that the appearance of only a couple of Floquet bands, and
momentum anisotropy is consistent with experimental observations.~\cite{Gedik13}

\section{Results in the presence of phonons}\label{phonons}
The above results for the time-averaged distribution
functions after a quench are exact and will not evolve in time.
However if we turn on the electron-phonon coupling, inelastic scattering will cause the distribution
functions to relax,
we now study how this happens, and what is the resulting steady-state.
We first briefly outline the derivation of
the kinetic or rate equation in the presence of phonons within the Floquet formalism
(see \cite{Hanggi2005} for general discussions).
Let $W(t)$ be the density matrix obeying
\begin{eqnarray}
\frac{dW(t)}{dt} = -i \left[H,W(t)\right]
\end{eqnarray}
It is convenient
to be in the interaction representation,
$W_{I}(t) = e^{i H_{\rm ph} t}U^{\dagger}_{\rm el}(t,0)W(t) U_{\rm el}(t,0)e^{- i H_{\rm ph} t}$.
To ${\cal O}(H_{c}^2)$, the density matrix obeys the following equation of motion
\begin{eqnarray}
&&\frac{dW_I}{dt}=-i\left[H_{c,I}(t),W_I(t_0)\right]\nonumber\\
&&-\int_{t_0}^tdt'\left[H_{c,I}(t),\left[H_{c,I}(t'),W_{I}(t')\right]\right]
\end{eqnarray}
where
$H_{c,I}$ is in the interaction representation.
We assume that at the initial time $t_0$, the electrons and phonons are uncoupled so that
$W(t_0) = W^{\rm el}_0(t_0)\otimes W^{\rm ph}(t_0)$, and that
initially the electrons are in the state $|\Psi(t)\rangle$ described in Section~\ref{quench}, while
the phonons are in thermal equilibrium at temperature $T$. This is justified because phonon
dynamics is much slower than electron dynamics. Thus the quench state of Section~\ref{quench}
can be achieved within
femto-second time-scales,~\cite{Gedik13} while,
the phonons do not affect the system until pico-second
time-scales.

Thus,
\begin{eqnarray}
W^{\rm el}_0(t)= |\Psi(t)\rangle\langle \Psi(t)|=\prod_kW^{\rm el}_{k,0}
\end{eqnarray} where
\begin{eqnarray}
\!\!W^{\rm el}_{k,0}(t)=\!\!\! \sum_{\alpha,\beta=\pm}e^{-i(\epsilon_{k \alpha}-\epsilon_{k \beta})t}|\phi_{k\alpha}(t)\rangle
\langle\phi_{k\beta}(t)|
\rho_{k,\alpha\beta}^{\rm quench}
\end{eqnarray}
with
\begin{eqnarray}
\rho_{k,\alpha\beta}^{\rm quench}= \langle \phi_{k\alpha}(0)|\psi_{{\rm in},k} \rangle\langle\psi_{{\rm in},k}
| \phi_{k\beta}(0)\rangle
\end{eqnarray}

Defining the electron reduced
density matrix as the one obtained from tracing over the phonons,
$W^{\rm el} = {\rm Tr}_{\rm ph}W$, and noting that $H_c$ being linear in the phonon operators,
the trace vanishes, we need to solve,
\begin{eqnarray}
\frac{dW^{\rm el}_I}{dt}=-{\rm Tr}_{\rm ph}\int_{t_0}^tdt'\left[H_{c,I}(t),\left[H_{c,I}(t'),W_{I}(t')\right]\right]
\end{eqnarray}
We assume that the phonons are an ideal reservoir and stay in equilibrium. In that case $W_I(t) = W^{\rm el}_I(t)\otimes
e^{-H_{\rm ph}/T}/{\rm Tr}\left[ e^{-H_{\rm ph}/T}\right]$ (we set $k_B=1$).

The most general form of the reduced density matrix
for the electrons is
\begin{eqnarray}
W^{\rm el}_I(t)  =\prod_k \sum_{\alpha\beta}\rho_{k,\alpha \beta}(t) |\phi_{k,\alpha}(t)\rangle\langle\phi_{k,\beta}(t)|
\end{eqnarray}
where in the absence of phonons, $\rho_{k,\alpha \beta}=\rho_{k,\alpha\beta}^{\rm quench }$ and are time-independent
in the interaction representation.
The last remaining assumption is to identify the slow
and fast variables, which allows one to make the Markov approximation.~\cite{Hanggi2005}
We write $\rho_{k,\alpha\beta}(t)= \sum_{m={\rm int}}e^{i m \Omega t}\rho_{k,\alpha\beta}^{(m)}(t)$
where in what follows we assume that $\rho_{k,\alpha \beta}^{(m)}(t)$ are slowly varying on time scales
of the period of the
AC field and the relevant phonon frequencies.
In addition we only study the diagonal components of $\rho_{k,\alpha\alpha}^{(m)}$, which after the Markov approximation,
obey the rate equation
\begin{eqnarray}
\!\!\!\!\!\!\!\left[\dot{\rho}_{k,\alpha\alpha}^{(m)}(t) + i m\Omega \rho_{k,\alpha\alpha}^{(m)}(t)\right]
=-\!\!\!\!\sum_{m',\beta=\pm}\!\!L^{m,m'}_{k,\alpha\beta}\!
\rho_{k,\beta\beta}^{(m-m')}(t)
\end{eqnarray}
The initial condition we will consider corresponds to a quench switch on protocol
for the laser $\rho_{k,\alpha\alpha}^{(m)}(t=0)=\delta_{m=0}
\rho_{k,\alpha\alpha}^{\rm quench}$, with the
in-scattering and out-scattering rates $L^{m,m'}_{k,\alpha\beta}$ given in Appendix~\ref{k0ph}.

Since the rate equation is a weak-coupling quasi-classical approximation in the
electron-phonon coupling, the position of the resonances in
the spectral density are not modified, and thus even with phonons,
$g^R$ is unchanged. The phonons strongly modify the steady-state lesser Green's function because the
distribution function of the electrons is changed due to inelastic scattering with phonons. In the numerical
solutions for the rate equation we assume optical phonons with a uniform phonon density of states $D_{\rm ph}$,
with a broad band-width so that inelastic scattering is always possible. We also assume 
an isotropic electron-phonon coupling $\lambda_x=\lambda_y=\lambda$. The results can easily be generalized to
optical phonons with narrow band-widths, as for frequencies below or above the optical phonon frequencies, the distribution function will
remain unchanged, and will be given by that for the quench.

The time-evolution of the density matrix from a quench-type initial state is shown in Fig.~\ref{fig5}, 
where the rate for reaching steady-state
is set by the strength of the electron-phonon coupling $\lambda^2D_{\rm ph}$.
In what follows, we present
results for $G^<$ at long times when a steady-state has been reached.
The solution of the rate
equations in Fig.~\ref{fig5} shows 
that the steady-state is characterized by some oscillations with time (controlled by $\lambda^2D_{\rm ph}$),
and our results are presented after a time-averaging of $\overline{\rho_{k,\alpha\alpha}(t)}=\rho_{k,\alpha\alpha}^{\rm ss}$
over several cycles. After this time-averaging, the steady-state
lesser Green's function in the presence of phonons is given by,
\begin{eqnarray}
&&G^{<}_{\sigma\sigma'}(k,t,t')=\nonumber\\
&&-i\sum_{\alpha_{=\pm}}\biggl[\rho_{k,\alpha\alpha}^{\rm ss}\langle\phi_{k,\alpha}(0)|c_{k\sigma}^{\dagger}(t)
c_{k\sigma'}(t')|\phi_{k,\alpha}(0)\rangle\biggr]
\end{eqnarray}
where $c_{k\sigma}(t) = \sum_{\sigma'}
U_{k\sigma\sigma'}(t,0)c_{k\sigma'}(0)$. Note that due to the laser field, $G^{<}$ is not time-translationally invariant,
and so we average over the mean time $(t+t')/2$ in a manner similar to that done in Section~\ref{quench}.

Remarkably, for $k=0$, $L^{m,m'}_{k=0,\alpha\beta}=\delta_{m,m'}L^m_{k=0,\alpha\beta}$,
so that again analytic results are possible. Here we find for the spin-density at $k=0$,
\begin{eqnarray}
&&P_z(k=0;H_c\neq 0)=\nonumber\\
&&\frac{-2\Omega\left(\Delta^2+\Omega^2\right)/\Delta}{\sum_{\alpha=\pm}\left(\Delta-\alpha\Omega\right)^2\biggl\{
1+2N\left(\Delta + \alpha \Omega\right)\biggr\}}\label{Pzph}
\end{eqnarray}
where $N(x)$ is the Bose distribution function, while
\begin{eqnarray}
&&iG^{<}_{\sigma\sigma}(k=0,\omega;H_c\neq 0) = 2\pi\sum_{\alpha=\pm}\rho_{k=0,\alpha\alpha}^{\rm ss} \nonumber\\
&&\times \biggl[
\left(\frac{\Delta +\alpha \sigma \Omega}{2\Delta}\right)
\delta\left(\omega +\sigma\frac{(\alpha\Delta + \Omega)}{2}\right) \biggr]\label{glessmainph}
\end{eqnarray}
where~\cite{Holthaus14}
\begin{eqnarray}
\rho_{k=0,++}^{\rm ss}=\frac{\sum_{\beta=\pm}\left(\Delta-\beta\Omega\right)^2
N\left(\Delta + \beta\Omega\right)}{\sum_{\alpha=\pm}\left(\Delta-\alpha\Omega\right)^2\left(
1+2N\left(\Delta + \alpha \Omega\right)\right)}
\end{eqnarray}
with $\sum_{\alpha=\pm}\rho_{k=0,\alpha\alpha}^{\rm ss}=1$.
Note that the above results at $k=0$ are isotropic in being independent of the angle $\theta_k$.
Thus the coupling to phonons makes the electrons lose memory of the initial state as well as the initial switch-on protocol
for the laser. This results in a symmetric
distribution of the density in momentum space. This is also clearly seen in the contour plot of
Fig.~\ref{fig1}. Fig.~\ref{fig2} shows that the spin-density still retains oscillations at momenta $k$ for which the
photon frequencies are resonant with the Dirac bands, however the magnitude of the oscillations decay with
increasing temperature of the phonon bath, with the spin-density $P_z(k)$ approaching zero as the temperature increases.
\begin{figure}
\centering
\includegraphics[height=10cm,width=8.75cm,keepaspectratio]{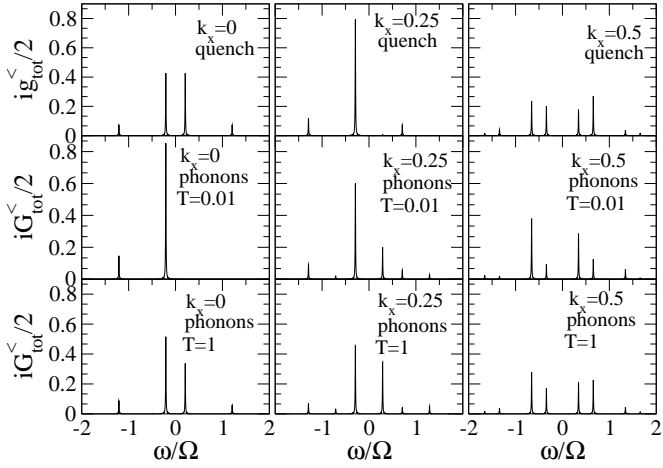}
\caption{$iG^{<}_{\rm tot}(k,\omega)$ for the quench with no phonons (upper panels) and at steady-state
with phonons at temperature $T=0.01\Omega,1\Omega$ (middle and lower panels) for $k_y=10^{-4}$ and $k_x=0.0,0.25,0.5$.
$A_0/\Omega=0.5,\lambda^2D_{\rm ph}=0.1\Omega,\Omega=1.0$. Normalization such that the peak heights equal the prefactor of
the $\delta$-functions.
}
\label{fig4}
\end{figure}

The spin averaged ARPES spectrum
$iG^<_{\rm tot}$ in the steady-state with phonons is shown as an intensity plot in the middle and
lower panels in Fig.~\ref{fig3} and along some momentum slices in Fig.~\ref{fig4}.
One finds that as the temperature of the phonon bath decreases, the magnitude of the
resonances at positive frequencies decrease and the ones at negative frequencies increase, maintaining the
sum rule. While this is also the expected result from a simple thermal Green's function where the
weights of the resonances are $\delta(\omega-\epsilon_k)n_F(\epsilon_k)$, yet note that the precise weights in
steady-state are not thermal. This can also be clearly seen in the analytic solution for $k=0$.
In particular Eq.~(\ref{Pzph})  implies that in the high frequency limit
\begin{eqnarray}
&&P_z(k=0,A_0\ll \Omega)\rightarrow\nonumber\\
&&\frac{\tanh\left[A_0^2/\Omega T\right]}{\left[1+(A_0^4/\Omega^4)\tanh\left(A_0^2/\Omega T\right)\coth\left(\Omega/T\right)\right]}
\end{eqnarray}
In this high-frequency limit, the Floquet Hamiltonian is $H_F\simeq \sigma_xk_x+\sigma_yk_y + \sigma_zA_0^2/\Omega$,
so a naive guess would be that the thermalized state should have a magnetization of $\tanh\left(h_z/2T\right)$
where $h_z=2A_0^2/\Omega$. The result for $P_z$ shows deviations from this guess at ${\cal O}(A_0^4/\Omega^4)$.
Thus, the presence of the AC drive causes the electrons to reach a nonequilibrium steady-state even when the
phonon reservoir to which the electrons are coupled are themselves always in thermal equilibrium.
Fig.~\ref{fig4} also shows that as $k$ approaches
the photon induced resonance condition $|k|\sim n\Omega/2$, the effective temperature is higher, as more frequencies are
excited. This result is clearly reflected in Fig.~\ref{fig3} (central panel) where even when the phonon temperature
is very low, the avoided crossings are characterized by a high population density.
\begin{figure}
\centering
\includegraphics[height=8cm,width=8cm,keepaspectratio]{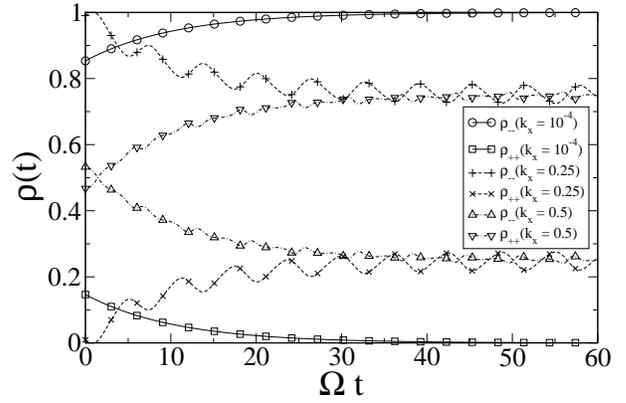}
\caption{Time-evolution of $\rho_{k,\alpha\alpha=\pm}$ from an initial state corresponding to a quench
for $k_y$=$0$ and $k_x=10^{-4},0.25,0.5$.
$A_0/\Omega=0.5,\lambda^2D_{\rm ph}=0.1\Omega,T=0.01\Omega,\Omega=1.0$.
}
\label{fig5}
\end{figure}

\section{Conclusions}\label{conclu}

In summary we have studied the electron distribution in a Floquet topological system under two circumstances, one is for
the closed system, where the results are very sensitive to how the AC field has been switched on,
showing highly anisotropic distribution functions,
the second is for the open system where the electrons are coupled to a reservoir of phonons.
While coupling
to phonons causes the system to lose memory of its initial state, yet the presence of the drive gives rise
to non-trivial nonequilibrium steady-states observable in ARPES. Since electron dynamics is much faster than
phonon dynamics, the results for the quench should be observable on short $\sim$ femto-second time-scales,
while the phonons will start relaxing the system on much longer time-scales. 
The rate equations show that
the system will eventually reach a nonequilibrium steady-state with the phonons on time-scales that are 
inverse of the effective electron-phonon coupling ${\cal O}(\lambda^2 D_{\rm ph})$ which is a highly material dependent parameter,
and in realistic materials suggests time-scales of the order of pico-seconds.

An important open question is to understand
transport phenomena such as the Hall conductance. Since the
Hall response is dominated by the behavior near $k=0$
where the Berry curvature is peaked, our results imply that
the anisotropic distribution of the closed system will cause significant deviation from the quantum
limit. On the other hand coupling to low temperature phonons
induces cooling of Floquet states.
The cooling works efficiently near the Dirac point, which could help the system to approach
the quantum limit. However, near resonant points (energy difference $\sim n\Omega$), our results also show that
the effective temperature stays high due to photo-carriers. Thus how close the system is to the quantum
limit will be a competition between the contribution to the Hall-conductance at the Dirac points,
and the role of these excited photo-carriers. A quantitative treatment requires
being on the lattice, as in the continuum, the Berry-curvature shows very sharp peaks at the resonances, 
and becomes mathematically ill-defined. 
However it is clear that to reach the quantum limit, in addition to having a low temperature bath, it would
also be helpful to be in the high frequency regime where $\Omega$ is greater than the band-width so as to
suppress excited photo-carriers, a regime which is non-existent in the continuum due to the unbounded
energy dispersion.

{\sl Acknowledgments:}
This work was supported by US Department of Energy (DOE-BES)
under Award No. DE-SC0010821 (HD and AM), and partially by the Simons Foundation (academic year support for AM).

\appendix

\section{Green's functions} \label{gf}
In this section we highlight how the Green's functions defined in Eqns.~(\ref{grdef}) and~(\ref{glessdef})
can be obtained.
Since the quasi-modes are periodic in time, we may write them as,
\begin{eqnarray}
|\phi_{k\alpha}(t) \rangle = \sum_{m \in {\rm int}} e^{im\Omega t}\begin{pmatrix}\alpha_{km\alpha}\\\beta_{km\alpha}\end{pmatrix}
\end{eqnarray}
Thus the time-evolution operator becomes,
\begin{eqnarray}
&&U_k(t=T_m+\tau/2,t'=T_m-\tau/2)= \nonumber\\
&&\sum_{\alpha=\pm,m,m'}e^{-i \epsilon_{k\alpha} \tau + \frac{m+m'}{2}\Omega\tau + i (m-m')\Omega T_m}\nonumber\\
&&\times \begin{pmatrix}\alpha_{km\alpha}\\\beta_{km\alpha}\end{pmatrix}\begin{pmatrix}\alpha_{km'\alpha}^*&\beta_{km'\alpha}^*\end{pmatrix}
\end{eqnarray}
Averaging over $T_m$,
\begin{eqnarray}
&&\overline{U}_k(\tau)= \nonumber\\
&&\sum_{\alpha=\pm, m}e^{-i \epsilon_{k\alpha} \tau + m\Omega\tau}
\begin{pmatrix}\alpha_{km\alpha}\\\beta_{km\alpha}\end{pmatrix}\begin{pmatrix}\alpha_{km\alpha}^*&\beta_{km\alpha}^*\end{pmatrix}
\end{eqnarray}
so that on Fourier transforming with respect to the time
difference $\tau$, the retarded Green's function becomes
\begin{eqnarray}
&&g^R(k,\omega)=\sum_{\alpha,m}\frac{1}{\omega -\left(\epsilon_{k\alpha}-m\Omega\right)+i\delta}\nonumber\\
&&\times \begin{pmatrix}\alpha_{km\alpha}\\\beta_{km\alpha}\end{pmatrix}\begin{pmatrix}\alpha_{km\alpha}^*&\beta_{km\alpha}^*\end{pmatrix}
\label{grgen}
\end{eqnarray}
For $k=0$, analytic expressions for $g_R$ may be obtained and these are presented in Eq.~(\ref{gRk0sol}),

For the lesser Green's functions, using Eq.~(\ref{glessdef}), we have,
\begin{eqnarray}
&&g^{<}_{\uparrow\uparrow}(k,t,t')=-i\sum_{\alpha,\beta=\pm,m,m',n,n'}e^{-i\epsilon_{k\beta}t'+i\epsilon_{k\alpha} t+i n \Omega t'-i m\Omega t}\nonumber\\
&&\biggl[
\alpha_{kn\beta}\alpha_{kn'\beta}^* \alpha_{km'\alpha}\alpha^*_{km\alpha}\langle c_{k\uparrow}^{\dagger}c_{k\uparrow}\rangle_0\nonumber\\
&&+\alpha_{kn\beta}\beta_{kn'\beta}^* \beta_{km'\alpha}\alpha^*_{km\alpha}\langle c_{k\downarrow}^{\dagger}c_{k\downarrow}\rangle_0\nonumber\\
&&  +
\alpha_{kn\beta}\alpha_{kn'\beta}^* \beta_{km'\alpha}\alpha^*_{km\alpha}\langle c_{k\downarrow}^{\dagger}c_{k\uparrow}\rangle_0\nonumber\\
&&+ \alpha_{kn\beta}\beta_{kn'\beta}^* \alpha_{km'\alpha}\alpha^*_{km\alpha}\langle c_{k\uparrow}^{\dagger}c_{k\downarrow}\rangle_0
\biggr]\nonumber\\\label{glessup}
\end{eqnarray}
and,
\begin{eqnarray}
&&g^{<}_{\downarrow\downarrow}(k,t,t')=-i\sum_{\alpha,\beta=\pm,m,m',n,n'}e^{-i\epsilon_{k\beta}t'+i\epsilon_{k\alpha} t+i n \Omega t'-i m\Omega t}\nonumber\\
&&\biggl[
\beta_{kn\beta}\alpha_{kn'\beta}^* \alpha_{km'\alpha}\beta^*_{km\alpha}\langle c_{k\uparrow}^{\dagger}c_{k\uparrow}\rangle_0\nonumber\\
&&+\beta_{kn\beta}\beta_{kn'\beta}^* \beta_{km'\alpha}\beta^*_{km\alpha}\langle c_{k\downarrow}^{\dagger}c_{k\downarrow}\rangle_0 \nonumber\\
&&+
\beta_{kn\beta}\alpha_{kn'\beta}^* \beta_{km'\alpha}\beta^*_{km\alpha}\langle c_{k\downarrow}^{\dagger}c_{k\uparrow}\rangle_0\nonumber\\
&&+ \beta_{kn\beta}\beta_{kn'\beta}^* \alpha_{km'\alpha}\beta^*_{km\alpha}\langle c_{k\uparrow}^{\dagger}c_{k\downarrow}\rangle_0
\biggr]\nonumber\\\label{glessdown}
\end{eqnarray}
where $\langle c_{k\sigma}^{\dagger}c_{k\sigma'}\rangle_0 = \langle\psi_{{\rm in},k}|c_{k\sigma}^{\dagger}c_{k\sigma'}|\psi_{{\rm in},k} \rangle$
for the closed system with a quench switch-on protocol. For the open system in steady-state, $\langle c_{k\sigma}^{\dagger}c_{k\sigma'}\rangle_0$
is the average with respect to the steady-state reduced density matrix of the electrons, which is in turn obtained from solving
a kinetic equation.

Time-averaging over the mean time $T_m=(t+t')/2$ imposes $\alpha=\beta, m=n$, so that
\begin{eqnarray}
&&\overline{g}^{<}_{\uparrow\uparrow}(k,\tau=t'-t)=-i\sum_{\alpha=\pm,n}e^{-i\left[\epsilon_{k\alpha}-n\Omega\right]\tau}|\alpha_{kn\alpha}|^2\rho_{k,\alpha\alpha}
\nonumber\\
&&\overline{g}^{<}_{\downarrow\downarrow}(k,\tau=t'-t)=-i\sum_{\alpha=\pm,n}e^{-i\left[\epsilon_{k\alpha}-n\Omega\right]\tau}|\beta_{kn\alpha}|^2\rho_{k,\alpha\alpha}
\nonumber\\
\end{eqnarray}
Above $\rho_{k,\alpha\alpha} = |\langle \phi_{k,\alpha}(0)|\psi_{{\rm in},k} \rangle|^2=\rho_{k,\alpha\alpha}^{\rm quench}$ 
for the quench in the closed system,
while it is obtained from a kinetic equation for the open system. 
For the latter, inelastic scattering causes $\rho_{k,\alpha\alpha}$ to evolve in time, 
and the Markov approximation that we will employ requires that this time-dependence is slow as compared to
all other time-scales. For the open system, we will then present results for the
Green's functions only at long times, where a steady-state has been reached, where the density matrix is
replaced by its steady-state value $\rho_{k,\alpha\alpha} = \rho_{k,\alpha\alpha}^{ss}$. Sometimes,
some slow residual oscillations 
such as those shown in Fig~\ref{fig5} persist even at long times, in this case
such slow oscillations will also be averaged over.

Fourier transforming,
\begin{eqnarray}
&&i{g}_{\uparrow\uparrow}^{<}(k,\omega) = 2\pi\sum_{n\alpha}\delta\left(\omega-\left[\epsilon_{k\alpha}-n\Omega\right]\right)|\alpha_{kn\alpha}|^2
\rho_{k,\alpha\alpha}\nonumber\\
&&i{g}_{\downarrow\downarrow}^{<}(k,\omega) = 2\pi\sum_{n\alpha}\delta\left(\omega-\left[\epsilon_{k\alpha}-n\Omega\right]\right)
|\beta_{kn\alpha}|^2\rho_{k,\alpha\alpha}\nonumber\\\label{gkgen}
\end{eqnarray}
Analytic expressions for the lesser Green's function for $k=0$ are given in Eq.~(\ref{glessmain}) 
for the quench and in Eq.~(\ref{glessmainph}) for
the steady-state with phonons.

\section{Analytic solution at $k=0$ for the quench (no phonons)}\label{k0}

Let us consider the solution of $H_{\rm el}$ when $k=0$.
In this case, the quasi-modes $|\phi_{\alpha}\rangle$ (we suppress the $k=0$ label) obey the equation,
\begin{eqnarray}
&&H_{\rm el,F}(k=0)|\phi_{\alpha}\rangle=\epsilon_{\alpha}|\phi_{\alpha}\rangle\\
&&H_{\rm el, F}(k=0)= \vec{A}\cdot\vec{\sigma}-i\partial_t\\
&&|\phi_{\alpha}\rangle= \begin{pmatrix}\phi_{\uparrow \alpha}\\\phi_{\downarrow \alpha}\end{pmatrix}
\end{eqnarray}
where $\vec{A}=A_0\left(\cos{\Omega t},-\sin{\Omega t}\right)$, so that
$\vec{A}\cdot\vec{\sigma}=A_0\begin{pmatrix}0&&e^{i\Omega t}\\e^{-i\Omega t}&& 0
\end{pmatrix} $.
Thus, the $\phi_{\uparrow,\downarrow \alpha}$ obey the coupled equation
\begin{eqnarray}
&&-i\partial_t \phi_{\uparrow \alpha}+A_0 e^{i\Omega t}\phi_{\downarrow \alpha}=\epsilon_{\alpha}\phi_{\uparrow \alpha}\\
&&-i\partial_t \phi_{\downarrow \alpha}+A_0 e^{-i\Omega t}\phi_{\uparrow \alpha}=\epsilon_{\alpha}\phi_{\downarrow \alpha}
\end{eqnarray}
Substituting for
\begin{eqnarray}
\phi_{\downarrow \alpha}=\frac{e^{-i\Omega t}}{A_0}\left[\epsilon_{\alpha}\phi_{\uparrow \alpha}+i\partial_t \phi_{\uparrow
\alpha}\right]\label{phi12}
\end{eqnarray}
into the second equation above gives,
\begin{eqnarray}
\partial_t^2\phi_{\uparrow \alpha}-i\left[2\epsilon_{\alpha}+\Omega\right]\partial_t\phi_{\uparrow \alpha}
+\left(A_0^2-\Omega \epsilon_{\alpha}
-\epsilon_{\alpha}^2\right)\phi_{\uparrow \alpha}=0\nonumber\\
\end{eqnarray}
Writing $\phi_{\uparrow \pm}=d_{\uparrow \pm}e^{i\lambda_{\pm}t}$, one obtains
$\lambda_{\mp}=\frac{\Omega}{2}+\epsilon_{\mp}\pm \frac{\Delta}{2}$
where
\begin{eqnarray}
\Delta =\sqrt{4 A_0^2 + \Omega^2}
\end{eqnarray}
Since $\phi_{\uparrow ,\downarrow \alpha}(t+T)=\phi_{\uparrow,\downarrow \alpha}(t)$, $\lambda = m \Omega$, where $m$
is an integer. Thus Eq.~(\ref{phi12}) gives,
\begin{eqnarray}
\phi_{\downarrow \alpha}= d_{\downarrow \alpha}e^{i(m-1)\Omega t}\,\,;\phi_{\uparrow \alpha}= d_{\uparrow \alpha}e^{im\Omega t}
\end{eqnarray}
with
\begin{eqnarray}
&&\epsilon_{\pm}= \left(m-\frac{1}{2}\right)\Omega \pm \frac{\Delta}{2}\\
&&\frac{d_{\downarrow \pm}}{d_{\uparrow\pm}}=\frac{-\Omega \pm \Delta}{2A_0}
\end{eqnarray}
Thus,
\begin{eqnarray}
d_{\uparrow \pm}= \frac{\sqrt{2}A_0}{\sqrt{\Delta \left(\Delta \mp \Omega\right)}}\,\,;d_{\downarrow \pm}
= \pm\frac{1}{\sqrt{2}}\sqrt{1\mp \frac{\Omega}{\Delta}}
\end{eqnarray}
\begin{eqnarray}
|\phi_{\pm}(t)\rangle= e^{im\Omega t}\begin{pmatrix}d_{\uparrow \pm}\\ e^{-i\Omega t}d_{\downarrow \pm}
\end{pmatrix}
\end{eqnarray}
Note that while there are infinite possible ways to choose the quasi-modes and the corresponding quasi-energies, where
the quasi-energies are related by shifts by integer multiples of the frequency $\Omega$, this degeneracy is absent
in the wavefunctions corresponding to the exact solutions of the Schr\"odinger equation,
$|\Psi_{\alpha}(t)\rangle= e^{-i\epsilon_\alpha t}|\phi_{\alpha}(t)\rangle$. In particular the wavefunctions
are
\begin{eqnarray}
|\Psi_+(t)\rangle = e^{i\Omega t/2 -i\Delta t/2}\begin{pmatrix}\frac{\sqrt{2}A_0}{\sqrt{\Delta \left(\Delta - \Omega\right)}} \\
e^{-i\Omega t}\frac{1}{\sqrt{2}}\sqrt{1- \frac{\Omega}{\Delta}}\end{pmatrix} \label{pp}\\
|\Psi_-(t)\rangle = e^{i\Omega t/2 + i \Delta t/2}\begin{pmatrix} \frac{\sqrt{2}A_0}{\sqrt{\Delta \left(\Delta + \Omega\right)}}\\
-e^{-i\Omega t}\frac{1}{\sqrt{2}}\sqrt{1+ \frac{\Omega}{\Delta}}
\end{pmatrix}\label{pm}
\end{eqnarray}
One may also construct the time-evolution operator,
\begin{eqnarray}
&&U_{k=0}(t,t')= \sum_{\alpha=\pm}e^{-i\epsilon_{\alpha}(t-t')}|\phi_{\alpha}(t)\rangle\langle\phi_{\alpha}(t')|\\
&&=\!\!\sum_{\alpha=\pm}e^{-i\left(\frac{-\Omega + \alpha\Delta}{2}\right)(t-t')}\begin{pmatrix}
d_{\uparrow \alpha}\\ e^{-i\Omega t}d_{\downarrow \alpha} \end{pmatrix}\begin{pmatrix}
d_{\uparrow \alpha}&e^{i\Omega t'}d_{\downarrow \alpha} \end{pmatrix}\nonumber\\
\end{eqnarray}

If the state just before switching on the AC field is the ground state of the Dirac fermions,
\begin{eqnarray}
|\psi_{\rm in}\rangle = \frac{1}{\sqrt{2}}\begin{pmatrix}-e^{-i\theta_k}\\1\end{pmatrix}
\end{eqnarray}
then the wavefunction after the sudden switch-on of the AC field is given by
\begin{eqnarray}
&&|\Psi(t)\rangle = U_{k=0}(t,0)|\psi_{\rm in}\rangle \nonumber\\
&&= \sum_{\alpha=\pm}C_{-\alpha}\frac{\sqrt{\Delta(\Delta-\alpha\Omega)}}{\sqrt{2}A_0}
|\Psi_{\alpha}(t)\rangle
\end{eqnarray}
where
\begin{eqnarray}
C_{+}=-\frac{\left[\omega_-\psi_{\uparrow}(0)+A_0\psi_{\downarrow}(0)\right]}{\left(\omega_+-\omega_-\right)}\\
C_{-}= \frac{\left[\omega_+\psi_{\uparrow}(0)+A_0\psi_{\downarrow}(0)\right]}{\left(\omega_+-\omega_-\right)}
\end{eqnarray}
with
\begin{eqnarray}
\omega_{\pm}=\frac{\Omega\pm \Delta}{2}\\
\psi_{\uparrow}(0) = -e^{-i\theta_k}/\sqrt{2}; \psi_{\downarrow}(0)=1/\sqrt{2}
\end{eqnarray}
Once the wavefunction $|\Psi(t)\rangle$ and the time evolution
operator $U(t,t')$ are known, one may compute all the single-time and two-time averages discussed in the main text.

Using the above, and Eq.~(\ref{grgen}), the expression for the retarded Green's function is,
\begin{eqnarray}
&&g^R(k=0,\omega)=\sum_{\alpha=\pm}\frac{1}{\omega -\left(-\frac{\Omega}{2}+\alpha\frac{\Delta}{2}\right)+i\delta}
\begin{pmatrix}d_{\uparrow \alpha}^2&0\\0&0\end{pmatrix} \nonumber\\
&&+ \sum_{\alpha=\pm}\frac{1}{\omega -\left(\frac{\Omega}{2}+\alpha\frac{\Delta}{2}\right)+i\delta}
\begin{pmatrix}0&0\\0&d_{\downarrow \alpha}^2\end{pmatrix}\label{gRk0sol}
\end{eqnarray}
Using Eq.~(\ref{gkgen}), the lesser Green's function is given in Eq.~(\ref{glessmain}).

\section{Rate equations for general $k$ and exact solution at $k=0$}\label{k0ph}
The rate equations after the Markov approximation are found to be (below $N_q=N(\omega_{q})$ is the Bose distribution function)
\begin{widetext}
\begin{eqnarray}
&&\left[\dot{\rho}_{k,\alpha \alpha}^{(m)}(t) + i m\Omega \rho_{k,\alpha\alpha}^{(m)}\right]=
 -\sum_{q,i=x,y,\beta=\pm,n_1,n_2}\left[\pi\lambda_{iq}^2\left(\epsilon^{i\bar{i}}\biggl\{C^{n_1}_{1k\alpha\beta}C^{n_2}_{1k\beta\alpha}+ C^{n_1}_{2k\alpha\beta}C^{n_2}_{2k\beta\alpha}\biggr\}
+C^{n_1}_{1k\alpha\beta}C^{n_2}_{2k\beta\alpha}
+C^{n_1}_{2k\alpha\beta}C^{n_2}_{1k\beta\alpha} \right) \right]\nonumber\\
&&\times \biggl[\biggl\{\left(1+N_q\right)
\delta(\epsilon_{k\beta}-\epsilon_{k\alpha}+(m-n_1)\Omega+\omega_{qi})+N_q\delta(\epsilon_{k\beta}-\epsilon_{k\alpha}+(m-n_1)\Omega-\omega_{qi})\biggr\}
\rho_{k,\alpha\alpha}^{(m-n_1-n_2)}(t)\nonumber \\
&&-\biggl\{\left(1+N_q\right)\delta(\epsilon_{k\beta}-\epsilon_{k\alpha}+(m-n_1)\Omega-\omega_{qi})+N_q\delta(\epsilon_{k\beta}-\epsilon_{k\alpha}+(m-n_1)\Omega+
\omega_{qi})\biggr\}
\rho_{k,\beta\beta}^{(m-n_1-n_2)}(t)\nonumber \\
&&+\biggl\{\left(1+N_q\right)
\delta(\epsilon_{k\beta}-\epsilon_{k\alpha}-(m-n_2)\Omega+\omega_{qi})+N_q\delta(\epsilon_{k\beta}-\epsilon_{k\alpha}-(m-n_2)\Omega-\omega_{qi})\biggr\}
\rho_{k,\alpha\alpha}^{(m-n_1-n_2)}(t)\nonumber \\
&&-\biggl\{\left(1+N_q\right)\delta(\epsilon_{k\beta}-\epsilon_{k\alpha}-(m-n_2)\Omega-\omega_{qi})+
N_q\delta(\epsilon_{k\beta}-\epsilon_{k\alpha}-(m-n_2)\Omega+\omega_{qi})\biggr\}
\rho_{k,\beta\beta}^{(m-n_1-n_2)}(t)
\biggr]\label{Rated3}
\end{eqnarray}
\end{widetext}
where $\epsilon^{x\bar{x}}=1,\epsilon^{y\bar{y}}=-1$, and
\begin{eqnarray}
\langle \phi_{k\alpha}(t)|c_{k\uparrow}^{\dagger}c_{k\downarrow}|\phi_{k\beta}(t)\rangle = \sum_{n}e^{in \Omega t}C_{1k\alpha\beta}^n\label{for1}\\
\langle \phi_{k\alpha}(t)|c_{k\downarrow}^{\dagger}c_{k\uparrow}|\phi_{k\beta}(t)\rangle = \sum_{n}e^{in \Omega t}C_{2k\alpha\beta}^n\label{for2}
\end{eqnarray}

\subsection{Analytic results for the rate equation at $k=0$}
At $k=0$, the exact expressions for the quasi-modes can be used to show that
\begin{eqnarray}
\langle \phi_{\alpha}(t)|c_{k=0,\uparrow}^{\dagger}c_{k=0,\downarrow}|\phi_{\beta}(t)\rangle=d_{\uparrow \alpha}d_{\downarrow \beta}e^{-i \Omega t}\\
\langle \phi_{\alpha}(t)|c_{k=0,\downarrow}^{\dagger}c_{k=0,\uparrow}|\phi_{\beta}(t)\rangle=d_{\downarrow\alpha}d_{\uparrow\beta}e^{i \Omega t}
\end{eqnarray}
Thus, the matrix elements entering the rate equation become,
\begin{eqnarray}
&&C^{(n)}_{1++}=\frac{A_0}{\Delta}\delta_{n=-1}; C^{(n)}_{1--}=-\frac{A_0}{\Delta}\delta_{n=-1}\nonumber\\
&&C^{(n)}_{1+-}=-\frac{1}{2}\left(1+\frac{\Omega}{\Delta}\right)\delta_{n=-1};
C^{(n)}_{1-+}=\frac{1}{2}\left(1-\frac{\Omega}{\Delta}\right)\delta_{n=-1}\nonumber\\
&&C^{(n)}_{2++}=\frac{A_0}{\Delta}\delta_{n=1}; C^{(n)}_{2--}=-\frac{A_0}{\Delta}\delta_{n=1}\nonumber\\
&&C^{(n)}_{2+-}=\frac{1}{2}\left(1-\frac{\Omega}{\Delta}\right)\delta_{n=1};
C^{(n)}_{2-+}=-\frac{1}{2}\left(1+\frac{\Omega}{\Delta}\right)\delta_{n=1}\nonumber
\end{eqnarray}
Let us assume $\lambda_{xq}=\lambda_{yq}$. In this case for $k=0$, $n_1+n_2=0$ in the rate equations. So for
$k=0$, the rate equations simplify to
\begin{eqnarray}
&&\partial_t\begin{pmatrix}\rho_{k=0,++}^{(m)}\\\rho_{k=0,--}^{(m)}\end{pmatrix}+
i m \Omega \begin{pmatrix}\rho_{k=0,++}^{(m)}\\\rho_{k=0,--}^{(m)}\end{pmatrix} \nonumber\\
&&= \begin{pmatrix}L^{(m)}_{k=0,++}&L^{(m)}_{k=0,+-}\\
L^{(m)}_{k=0,-+}&L^{(m)}_{k=0,--}\end{pmatrix}\begin{pmatrix}\rho^{(m)}_{k=0,++}\\\rho^{(m)}_{k=0,--}\end{pmatrix}
\end{eqnarray}
where $L^{(m)}_{k=0,++}=-L^{(m)}_{k=0,-+}; L^{(m)}_{k=0,+-}=-L^{(m)}_{k=0,--}$. We now make the assumption of
a uniform phonon density of states $D_{\rm ph}$ so that the rates are,
\begin{eqnarray}
&&L_{k=0,++}^{(m)}=-\pi\lambda^2D_{\rm ph}\frac{1}{2}\left(1+\frac{\Omega}{\Delta}\right)^2
\biggl[\nonumber\\
&&\{1+N(\Delta-\Omega - m\Omega)\}\theta(\Delta-\Omega - m\Omega)\nonumber\\
&&+
\{1+N(\Delta-\Omega + m\Omega)\}\theta(\Delta-\Omega + m\Omega)\nonumber\\
&& +N(-\Delta+\Omega - m\Omega)\theta(-\Delta+\Omega - m\Omega)\nonumber\\
&&+N(-\Delta+\Omega + m\Omega)\theta(-\Delta+\Omega + m\Omega)
\biggr]\nonumber \\
&&-\pi\lambda^2D_{\rm ph} \frac{1}{2}\left(1-\frac{\Omega}{\Delta}\right)^2\biggl[\nonumber\\
&&\{1+N(\Delta+\Omega - m\Omega)\}\theta(\Delta+\Omega - m\Omega)\nonumber\\
&&+ \{1+N(\Delta+\Omega + m\Omega)\}\theta(\Delta+\Omega + m\Omega)  \nonumber\\
&& +N(-\Delta-\Omega - m\Omega)\theta(-\Delta-\Omega - m\Omega)\nonumber\\
&&+N(-\Delta-\Omega + m\Omega)\theta(-\Delta-\Omega + m\Omega) \biggr]
\end{eqnarray}
and,
\begin{eqnarray}
&&L_{k=0,--}^{(m)}=-\pi\lambda^2D_{\rm ph}\frac{1}{2}\left(1-\frac{\Omega}{\Delta}\right)^2
\biggl[\nonumber\\
&&\{1+N(-\Delta-\Omega - m\Omega)\}\theta(-\Delta-\Omega - m\Omega)\nonumber\\
&&+ \{1+N(-\Delta-\Omega + m\Omega)\}\theta(-\Delta-\Omega + m\Omega) \nonumber\\
&&+N(\Delta+\Omega - m\Omega)\theta(\Delta+\Omega - m\Omega)\nonumber\\
&&+N(\Delta+\Omega + m\Omega)\theta(\Delta+\Omega + m\Omega) \biggr]\nonumber \\
&&-\pi\lambda^2D_{\rm ph} \frac{1}{2}\left(1+\frac{\Omega}{\Delta}\right)^2\biggl[\nonumber\\
&&\{1+N(-\Delta+\Omega - m\Omega)\}\theta(-\Delta+\Omega - m\Omega)\nonumber\\
&&+\{1+N(-\Delta+\Omega + m\Omega)\}\theta(-\Delta+\Omega + m\Omega) \nonumber\\
&&+N(\Delta-\Omega - m\Omega)\theta(\Delta-\Omega - m\Omega)\nonumber\\
&&+N(\Delta-\Omega + m\Omega)\theta(\Delta-\Omega + m\Omega) \biggr]
\end{eqnarray}
Above $\theta$ is the Heaviside step function.

%

%\bibliography{quench}

\begin{thebibliography}{41}%
\makeatletter
\providecommand \@ifxundefined [1]{%
 \@ifx{#1\undefined}
}%
\providecommand \@ifnum [1]{%
 \ifnum #1\expandafter \@firstoftwo
 \else \expandafter \@secondoftwo
 \fi
}%
\providecommand \@ifx [1]{%
 \ifx #1\expandafter \@firstoftwo
 \else \expandafter \@secondoftwo
 \fi
}%
\providecommand \natexlab [1]{#1}%
\providecommand \enquote  [1]{``#1''}%
\providecommand \bibnamefont  [1]{#1}%
\providecommand \bibfnamefont [1]{#1}%
\providecommand \citenamefont [1]{#1}%
\providecommand \href@noop [0]{\@secondoftwo}%
\providecommand \href [0]{\begingroup \@sanitize@url \@href}%
\providecommand \@href[1]{\@@startlink{#1}\@@href}%
\providecommand \@@href[1]{\endgroup#1\@@endlink}%
\providecommand \@sanitize@url [0]{\catcode `\\12\catcode `\$12\catcode
  `\&12\catcode `\#12\catcode `\^12\catcode `\_12\catcode `\%12\relax}%
\providecommand \@@startlink[1]{}%
\providecommand \@@endlink[0]{}%
\providecommand \url  [0]{\begingroup\@sanitize@url \@url }%
\providecommand \@url [1]{\endgroup\@href {#1}{\urlprefix }}%
\providecommand \urlprefix  [0]{URL }%
\providecommand \Eprint [0]{\href }%
\providecommand \doibase [0]{http://dx.doi.org/}%
\providecommand \selectlanguage [0]{\@gobble}%
\providecommand \bibinfo  [0]{\@secondoftwo}%
\providecommand \bibfield  [0]{\@secondoftwo}%
\providecommand \translation [1]{[#1]}%
\providecommand \BibitemOpen [0]{}%
\providecommand \bibitemStop [0]{}%
\providecommand \bibitemNoStop [0]{.\EOS\space}%
\providecommand \EOS [0]{\spacefactor3000\relax}%
\providecommand \BibitemShut  [1]{\csname bibitem#1\endcsname}%
\let\auto@bib@innerbib\@empty
%</preamble>
\bibitem [{\citenamefont {Haldane}(1988)}]{Haldane88}%
  \BibitemOpen
  \bibfield  {author} {\bibinfo {author} {\bibfnamefont {F.~D.~M.}\
  \bibnamefont {Haldane}},\ }\href {\doibase 10.1103/PhysRevLett.61.2015}
  {\bibfield  {journal} {\bibinfo  {journal} {Phys. Rev. Lett.}\ }\textbf
  {\bibinfo {volume} {61}},\ \bibinfo {pages} {2015} (\bibinfo {year}
  {1988})}\BibitemShut {NoStop}%
\bibitem [{\citenamefont {Hasan}\ and\ \citenamefont {Kane}(2010)}]{Hasan10}%
  \BibitemOpen
  \bibfield  {author} {\bibinfo {author} {\bibfnamefont {M.~Z.}\ \bibnamefont
  {Hasan}}\ and\ \bibinfo {author} {\bibfnamefont {C.~L.}\ \bibnamefont
  {Kane}},\ }\href {\doibase 10.1103/RevModPhys.82.3045} {\bibfield  {journal}
  {\bibinfo  {journal} {Rev. Mod. Phys.}\ }\textbf {\bibinfo {volume} {82}},\
  \bibinfo {pages} {3045} (\bibinfo {year} {2010})}\BibitemShut {NoStop}%
\bibitem [{\citenamefont {Qi}\ and\ \citenamefont {Zhang}(2011)}]{Zhang11}%
  \BibitemOpen
  \bibfield  {author} {\bibinfo {author} {\bibfnamefont {X.-L.}\ \bibnamefont
  {Qi}}\ and\ \bibinfo {author} {\bibfnamefont {S.-C.}\ \bibnamefont {Zhang}},\
  }\href {\doibase 10.1103/RevModPhys.83.1057} {\bibfield  {journal} {\bibinfo
  {journal} {Rev. Mod. Phys.}\ }\textbf {\bibinfo {volume} {83}},\ \bibinfo
  {pages} {1057} (\bibinfo {year} {2011})}\BibitemShut {NoStop}%
\bibitem [{\citenamefont {Kane}\ and\ \citenamefont {Mele}(2005)}]{Kane05}%
  \BibitemOpen
  \bibfield  {author} {\bibinfo {author} {\bibfnamefont {C.~L.}\ \bibnamefont
  {Kane}}\ and\ \bibinfo {author} {\bibfnamefont {E.~J.}\ \bibnamefont
  {Mele}},\ }\href {\doibase 10.1103/PhysRevLett.95.146802} {\bibfield
  {journal} {\bibinfo  {journal} {Phys. Rev. Lett.}\ }\textbf {\bibinfo
  {volume} {95}},\ \bibinfo {pages} {146802} (\bibinfo {year}
  {2005})}\BibitemShut {NoStop}%
\bibitem [{\citenamefont {Bernevig}\ \emph {et~al.}(2006)\citenamefont
  {Bernevig}, \citenamefont {Hughes},\ and\ \citenamefont
  {Zhang}}]{Bernevig06}%
  \BibitemOpen
  \bibfield  {author} {\bibinfo {author} {\bibfnamefont {B.~A.}\ \bibnamefont
  {Bernevig}}, \bibinfo {author} {\bibfnamefont {T.~L.}\ \bibnamefont
  {Hughes}}, \ and\ \bibinfo {author} {\bibfnamefont {S.-C.}\ \bibnamefont
  {Zhang}},\ }\href {\doibase 10.1126/science.1133734} {\bibfield  {journal}
  {\bibinfo  {journal} {Science}\ }\textbf {\bibinfo {volume} {314}},\ \bibinfo
  {pages} {1757} (\bibinfo {year} {2006})}\BibitemShut {NoStop}%
\bibitem [{\citenamefont {Senthil}(shed)}]{Senthil14}%
  \BibitemOpen
  \bibfield  {author} {\bibinfo {author} {\bibfnamefont {T.}~\bibnamefont
  {Senthil}},\ }\href@noop {} {\bibfield  {journal} {\bibinfo  {journal}
  {arXiv:1405.4015}\ } (\bibinfo {year} {unpublished})}\BibitemShut {NoStop}%
\bibitem [{\citenamefont {Oka}\ and\ \citenamefont {Aoki}(2009)}]{Oka09}%
  \BibitemOpen
  \bibfield  {author} {\bibinfo {author} {\bibfnamefont {T.}~\bibnamefont
  {Oka}}\ and\ \bibinfo {author} {\bibfnamefont {H.}~\bibnamefont {Aoki}},\
  }\href {\doibase 10.1103/PhysRevB.79.081406} {\bibfield  {journal} {\bibinfo
  {journal} {Phys. Rev. B}\ }\textbf {\bibinfo {volume} {79}},\ \bibinfo
  {pages} {081406} (\bibinfo {year} {2009})}\BibitemShut {NoStop}%
\bibitem [{\citenamefont {Inoue}\ and\ \citenamefont {Tanaka}(2010)}]{Inoue10}%
  \BibitemOpen
  \bibfield  {author} {\bibinfo {author} {\bibfnamefont {J.}~\bibnamefont
  {Inoue}}\ and\ \bibinfo {author} {\bibfnamefont {A.}~\bibnamefont {Tanaka}},\
  }\href {\doibase 10.1103/PhysRevLett.105.017401} {\bibfield  {journal}
  {\bibinfo  {journal} {Phys. Rev. Lett.}\ }\textbf {\bibinfo {volume} {105}},\
  \bibinfo {pages} {017401} (\bibinfo {year} {2010})}\BibitemShut {NoStop}%
\bibitem [{\citenamefont {Kitagawa}\ \emph {et~al.}(2010)\citenamefont
  {Kitagawa}, \citenamefont {Berg}, \citenamefont {Rudner},\ and\ \citenamefont
  {Demler}}]{Kitagawa10}%
  \BibitemOpen
  \bibfield  {author} {\bibinfo {author} {\bibfnamefont {T.}~\bibnamefont
  {Kitagawa}}, \bibinfo {author} {\bibfnamefont {E.}~\bibnamefont {Berg}},
  \bibinfo {author} {\bibfnamefont {M.}~\bibnamefont {Rudner}}, \ and\ \bibinfo
  {author} {\bibfnamefont {E.}~\bibnamefont {Demler}},\ }\href {\doibase
  10.1103/PhysRevB.82.235114} {\bibfield  {journal} {\bibinfo  {journal} {Phys.
  Rev. B}\ }\textbf {\bibinfo {volume} {82}},\ \bibinfo {pages} {235114}
  (\bibinfo {year} {2010})}\BibitemShut {NoStop}%
\bibitem [{\citenamefont {Lindner}\ \emph {et~al.}(2011)\citenamefont
  {Lindner}, \citenamefont {Refael},\ and\ \citenamefont
  {Galitski}}]{Lindner11}%
  \BibitemOpen
  \bibfield  {author} {\bibinfo {author} {\bibfnamefont {N.~H.}\ \bibnamefont
  {Lindner}}, \bibinfo {author} {\bibfnamefont {G.}~\bibnamefont {Refael}}, \
  and\ \bibinfo {author} {\bibfnamefont {V.}~\bibnamefont {Galitski}},\
  }\href@noop {} {\bibfield  {journal} {\bibinfo  {journal} {Nature Physics}\
  }\textbf {\bibinfo {volume} {7}},\ \bibinfo {pages} {490} (\bibinfo {year}
  {2011})}\BibitemShut {NoStop}%
\bibitem [{\citenamefont {Kitagawa}\ \emph {et~al.}(2011)\citenamefont
  {Kitagawa}, \citenamefont {Oka}, \citenamefont {Brataas}, \citenamefont
  {Fu},\ and\ \citenamefont {Demler}}]{Kitagawa11}%
  \BibitemOpen
  \bibfield  {author} {\bibinfo {author} {\bibfnamefont {T.}~\bibnamefont
  {Kitagawa}}, \bibinfo {author} {\bibfnamefont {T.}~\bibnamefont {Oka}},
  \bibinfo {author} {\bibfnamefont {A.}~\bibnamefont {Brataas}}, \bibinfo
  {author} {\bibfnamefont {L.}~\bibnamefont {Fu}}, \ and\ \bibinfo {author}
  {\bibfnamefont {E.}~\bibnamefont {Demler}},\ }\href {\doibase
  10.1103/PhysRevB.84.235108} {\bibfield  {journal} {\bibinfo  {journal} {Phys.
  Rev. B}\ }\textbf {\bibinfo {volume} {84}},\ \bibinfo {pages} {235108}
  (\bibinfo {year} {2011})}\BibitemShut {NoStop}%
\bibitem [{\citenamefont {Lindner}\ \emph {et~al.}(2013)\citenamefont
  {Lindner}, \citenamefont {Bergman}, \citenamefont {Refael},\ and\
  \citenamefont {Galitski}}]{Lindner13}%
  \BibitemOpen
  \bibfield  {author} {\bibinfo {author} {\bibfnamefont {N.~H.}\ \bibnamefont
  {Lindner}}, \bibinfo {author} {\bibfnamefont {D.~L.}\ \bibnamefont
  {Bergman}}, \bibinfo {author} {\bibfnamefont {G.}~\bibnamefont {Refael}}, \
  and\ \bibinfo {author} {\bibfnamefont {V.}~\bibnamefont {Galitski}},\ }\href
  {\doibase 10.1103/PhysRevB.87.235131} {\bibfield  {journal} {\bibinfo
  {journal} {Phys. Rev. B}\ }\textbf {\bibinfo {volume} {87}},\ \bibinfo
  {pages} {235131} (\bibinfo {year} {2013})}\BibitemShut {NoStop}%
\bibitem [{\citenamefont {G\'omez-Le\'on}\ and\ \citenamefont
  {Platero}(2013)}]{Gomez13}%
  \BibitemOpen
  \bibfield  {author} {\bibinfo {author} {\bibfnamefont {A.}~\bibnamefont
  {G\'omez-Le\'on}}\ and\ \bibinfo {author} {\bibfnamefont {G.}~\bibnamefont
  {Platero}},\ }\href {\doibase 10.1103/PhysRevLett.110.200403} {\bibfield
  {journal} {\bibinfo  {journal} {Phys. Rev. Lett.}\ }\textbf {\bibinfo
  {volume} {110}},\ \bibinfo {pages} {200403} (\bibinfo {year}
  {2013})}\BibitemShut {NoStop}%
\bibitem [{\citenamefont {Katan}\ and\ \citenamefont
  {Podolsky}(2013)}]{Podolsky13}%
  \BibitemOpen
  \bibfield  {author} {\bibinfo {author} {\bibfnamefont {Y.~T.}\ \bibnamefont
  {Katan}}\ and\ \bibinfo {author} {\bibfnamefont {D.}~\bibnamefont
  {Podolsky}},\ }\href {\doibase 10.1103/PhysRevLett.110.016802} {\bibfield
  {journal} {\bibinfo  {journal} {Phys. Rev. Lett.}\ }\textbf {\bibinfo
  {volume} {110}},\ \bibinfo {pages} {016802} (\bibinfo {year}
  {2013})}\BibitemShut {NoStop}%
\bibitem [{\citenamefont {Perez-Piskunow}\ \emph {et~al.}(2014)\citenamefont
  {Perez-Piskunow}, \citenamefont {Usaj}, \citenamefont {Balseiro},\ and\
  \citenamefont {Torres}}]{Usaj14}%
  \BibitemOpen
  \bibfield  {author} {\bibinfo {author} {\bibfnamefont {P.~M.}\ \bibnamefont
  {Perez-Piskunow}}, \bibinfo {author} {\bibfnamefont {G.}~\bibnamefont
  {Usaj}}, \bibinfo {author} {\bibfnamefont {C.~A.}\ \bibnamefont {Balseiro}},
  \ and\ \bibinfo {author} {\bibfnamefont {L.~E. F.~F.}\ \bibnamefont
  {Torres}},\ }\href {\doibase 10.1103/PhysRevB.89.121401} {\bibfield
  {journal} {\bibinfo  {journal} {Phys. Rev. B}\ }\textbf {\bibinfo {volume}
  {89}},\ \bibinfo {pages} {121401} (\bibinfo {year} {2014})}\BibitemShut
  {NoStop}%
\bibitem [{\citenamefont {Shirley}(1965)}]{Shirley65}%
  \BibitemOpen
  \bibfield  {author} {\bibinfo {author} {\bibfnamefont {J.~H.}\ \bibnamefont
  {Shirley}},\ }\href {\doibase 10.1103/PhysRev.138.B979} {\bibfield  {journal}
  {\bibinfo  {journal} {Phys. Rev.}\ }\textbf {\bibinfo {volume} {138}},\
  \bibinfo {pages} {B979} (\bibinfo {year} {1965})}\BibitemShut {NoStop}%
\bibitem [{\citenamefont {Sambe}(1973)}]{Sambe73}%
  \BibitemOpen
  \bibfield  {author} {\bibinfo {author} {\bibfnamefont {H.}~\bibnamefont
  {Sambe}},\ }\href {\doibase 10.1103/PhysRevA.7.2203} {\bibfield  {journal}
  {\bibinfo  {journal} {Phys. Rev. A}\ }\textbf {\bibinfo {volume} {7}},\
  \bibinfo {pages} {2203} (\bibinfo {year} {1973})}\BibitemShut {NoStop}%
\bibitem [{\citenamefont {Rudner}\ \emph {et~al.}(2013)\citenamefont {Rudner},
  \citenamefont {Lindner}, \citenamefont {Berg},\ and\ \citenamefont
  {Levin}}]{Rudner13}%
  \BibitemOpen
  \bibfield  {author} {\bibinfo {author} {\bibfnamefont {M.~S.}\ \bibnamefont
  {Rudner}}, \bibinfo {author} {\bibfnamefont {N.~H.}\ \bibnamefont {Lindner}},
  \bibinfo {author} {\bibfnamefont {E.}~\bibnamefont {Berg}}, \ and\ \bibinfo
  {author} {\bibfnamefont {M.}~\bibnamefont {Levin}},\ }\href {\doibase
  10.1103/PhysRevX.3.031005} {\bibfield  {journal} {\bibinfo  {journal} {Phys.
  Rev. X}\ }\textbf {\bibinfo {volume} {3}},\ \bibinfo {pages} {031005}
  (\bibinfo {year} {2013})}\BibitemShut {NoStop}%
\bibitem [{\citenamefont {Lababidi}\ \emph {et~al.}(2014)\citenamefont
  {Lababidi}, \citenamefont {Satija},\ and\ \citenamefont {Zhao}}]{Erhai14}%
  \BibitemOpen
  \bibfield  {author} {\bibinfo {author} {\bibfnamefont {M.}~\bibnamefont
  {Lababidi}}, \bibinfo {author} {\bibfnamefont {I.~I.}\ \bibnamefont
  {Satija}}, \ and\ \bibinfo {author} {\bibfnamefont {E.}~\bibnamefont
  {Zhao}},\ }\href {\doibase 10.1103/PhysRevLett.112.026805} {\bibfield
  {journal} {\bibinfo  {journal} {Phys. Rev. Lett.}\ }\textbf {\bibinfo
  {volume} {112}},\ \bibinfo {pages} {026805} (\bibinfo {year}
  {2014})}\BibitemShut {NoStop}%
\bibitem [{\citenamefont {Kundu}\ \emph {et~al.}(shed)\citenamefont {Kundu},
  \citenamefont {Fertig},\ and\ \citenamefont {Seradjeh}}]{Kundu14}%
  \BibitemOpen
  \bibfield  {author} {\bibinfo {author} {\bibfnamefont {A.}~\bibnamefont
  {Kundu}}, \bibinfo {author} {\bibfnamefont {H.}~\bibnamefont {Fertig}}, \
  and\ \bibinfo {author} {\bibfnamefont {B.}~\bibnamefont {Seradjeh}},\
  }\href@noop {} {\bibfield  {journal} {\bibinfo  {journal} {arXiv:1406.1490}\
  } (\bibinfo {year} {unpublished})}\BibitemShut {NoStop}%
\bibitem [{\citenamefont {Rechstman}\ \emph {et~al.}(2013)\citenamefont
  {Rechstman}, \citenamefont {Zeuner}, \citenamefont {Plotnik}, \citenamefont
  {Lumer}, \citenamefont {Podolsky}, \citenamefont {Dreisow}, \citenamefont
  {Nolte}, \citenamefont {Segev},\ and\ \citenamefont {Szameit}}]{Segev13}%
  \BibitemOpen
  \bibfield  {author} {\bibinfo {author} {\bibfnamefont {M.}~\bibnamefont
  {Rechstman}}, \bibinfo {author} {\bibfnamefont {J.}~\bibnamefont {Zeuner}},
  \bibinfo {author} {\bibfnamefont {Y.}~\bibnamefont {Plotnik}}, \bibinfo
  {author} {\bibfnamefont {Y.}~\bibnamefont {Lumer}}, \bibinfo {author}
  {\bibfnamefont {D.}~\bibnamefont {Podolsky}}, \bibinfo {author}
  {\bibfnamefont {F.}~\bibnamefont {Dreisow}}, \bibinfo {author} {\bibfnamefont
  {S.}~\bibnamefont {Nolte}}, \bibinfo {author} {\bibfnamefont
  {M.}~\bibnamefont {Segev}}, \ and\ \bibinfo {author} {\bibfnamefont
  {A.}~\bibnamefont {Szameit}},\ }\href@noop {} {\bibfield  {journal} {\bibinfo
   {journal} {Nature (London)}\ }\textbf {\bibinfo {volume} {496}},\ \bibinfo
  {pages} {196} (\bibinfo {year} {2013})}\BibitemShut {NoStop}%
\bibitem [{\citenamefont {Jotzu}\ \emph {et~al.}(shed)\citenamefont {Jotzu},
  \citenamefont {Messer}, \citenamefont {Desbuquois}, \citenamefont {Lebrat},
  \citenamefont {Uehlinger}, \citenamefont {Greif},\ and\ \citenamefont
  {Esslinger}}]{Esslinger14}%
  \BibitemOpen
  \bibfield  {author} {\bibinfo {author} {\bibfnamefont {G.}~\bibnamefont
  {Jotzu}}, \bibinfo {author} {\bibfnamefont {M.}~\bibnamefont {Messer}},
  \bibinfo {author} {\bibfnamefont {R.}~\bibnamefont {Desbuquois}}, \bibinfo
  {author} {\bibfnamefont {M.}~\bibnamefont {Lebrat}}, \bibinfo {author}
  {\bibfnamefont {T.}~\bibnamefont {Uehlinger}}, \bibinfo {author}
  {\bibfnamefont {D.}~\bibnamefont {Greif}}, \ and\ \bibinfo {author}
  {\bibfnamefont {T.}~\bibnamefont {Esslinger}},\ }\href@noop {} {\bibfield
  {journal} {\bibinfo  {journal} {arXiv:1406.7874}\ } (\bibinfo {year}
  {unpublished})}\BibitemShut {NoStop}%
\bibitem [{\citenamefont {Lazarides}\ \emph {et~al.}(2014)\citenamefont
  {Lazarides}, \citenamefont {Das},\ and\ \citenamefont
  {Moessner}}]{Lazarides14}%
  \BibitemOpen
  \bibfield  {author} {\bibinfo {author} {\bibfnamefont {A.}~\bibnamefont
  {Lazarides}}, \bibinfo {author} {\bibfnamefont {A.}~\bibnamefont {Das}}, \
  and\ \bibinfo {author} {\bibfnamefont {R.}~\bibnamefont {Moessner}},\ }\href
  {\doibase 10.1103/PhysRevLett.112.150401} {\bibfield  {journal} {\bibinfo
  {journal} {Phys. Rev. Lett.}\ }\textbf {\bibinfo {volume} {112}},\ \bibinfo
  {pages} {150401} (\bibinfo {year} {2014})}\BibitemShut {NoStop}%
\bibitem [{\citenamefont {Sentef}\ \emph {et~al.}(shed)\citenamefont {Sentef},
  \citenamefont {Claassen}, \citenamefont {Kemper}, \citenamefont {Moritz},
  \citenamefont {Oka}, \citenamefont {Freericks},\ and\ \citenamefont
  {Devereaux}}]{Sentef14}%
  \BibitemOpen
  \bibfield  {author} {\bibinfo {author} {\bibfnamefont {M.~A.}\ \bibnamefont
  {Sentef}}, \bibinfo {author} {\bibfnamefont {M.}~\bibnamefont {Claassen}},
  \bibinfo {author} {\bibfnamefont {A.~F.}\ \bibnamefont {Kemper}}, \bibinfo
  {author} {\bibfnamefont {B.}~\bibnamefont {Moritz}}, \bibinfo {author}
  {\bibfnamefont {T.}~\bibnamefont {Oka}}, \bibinfo {author} {\bibfnamefont
  {J.~K.}\ \bibnamefont {Freericks}}, \ and\ \bibinfo {author} {\bibfnamefont
  {T.~P.}\ \bibnamefont {Devereaux}},\ }\href@noop {} {\bibfield  {journal}
  {\bibinfo  {journal} {arXiv:1401.5103}\ } (\bibinfo {year}
  {unpublished})}\BibitemShut {NoStop}%
\bibitem [{\citenamefont {Goldman}\ and\ \citenamefont
  {Dalibard}(2014)}]{Dalibard14}%
  \BibitemOpen
  \bibfield  {author} {\bibinfo {author} {\bibfnamefont {N.}~\bibnamefont
  {Goldman}}\ and\ \bibinfo {author} {\bibfnamefont {J.}~\bibnamefont
  {Dalibard}},\ }\href@noop {} {\bibfield  {journal} {\bibinfo  {journal}
  {Phys. Rev. X}\ }\textbf {\bibinfo {volume} {4}},\ \bibinfo {pages} {031027}
  (\bibinfo {year} {2014})}\BibitemShut {NoStop}%
\bibitem [{\citenamefont {Gu}\ \emph {et~al.}(2011)\citenamefont {Gu},
  \citenamefont {Fertig}, \citenamefont {Arovas},\ and\ \citenamefont
  {Auerbach}}]{Fertig11}%
  \BibitemOpen
  \bibfield  {author} {\bibinfo {author} {\bibfnamefont {Z.}~\bibnamefont
  {Gu}}, \bibinfo {author} {\bibfnamefont {H.~A.}\ \bibnamefont {Fertig}},
  \bibinfo {author} {\bibfnamefont {D.~P.}\ \bibnamefont {Arovas}}, \ and\
  \bibinfo {author} {\bibfnamefont {A.}~\bibnamefont {Auerbach}},\ }\href
  {\doibase 10.1103/PhysRevLett.107.216601} {\bibfield  {journal} {\bibinfo
  {journal} {Phys. Rev. Lett.}\ }\textbf {\bibinfo {volume} {107}},\ \bibinfo
  {pages} {216601} (\bibinfo {year} {2011})}\BibitemShut {NoStop}%
\bibitem [{\citenamefont {Kundu}\ and\ \citenamefont
  {Seradjeh}(2013)}]{Kundu13}%
  \BibitemOpen
  \bibfield  {author} {\bibinfo {author} {\bibfnamefont {A.}~\bibnamefont
  {Kundu}}\ and\ \bibinfo {author} {\bibfnamefont {B.}~\bibnamefont
  {Seradjeh}},\ }\href {\doibase 10.1103/PhysRevLett.111.136402} {\bibfield
  {journal} {\bibinfo  {journal} {Phys. Rev. Lett.}\ }\textbf {\bibinfo
  {volume} {111}},\ \bibinfo {pages} {136402} (\bibinfo {year}
  {2013})}\BibitemShut {NoStop}%
\bibitem [{\citenamefont {Torres}\ \emph {et~al.}(shed)\citenamefont {Torres},
  \citenamefont {Perez-Piskunow}, \citenamefont {Balseiro},\ and\ \citenamefont
  {Usaj}}]{Torres14}%
  \BibitemOpen
  \bibfield  {author} {\bibinfo {author} {\bibfnamefont {L.~E. F.~F.}\
  \bibnamefont {Torres}}, \bibinfo {author} {\bibfnamefont {P.~M.}\
  \bibnamefont {Perez-Piskunow}}, \bibinfo {author} {\bibfnamefont {C.~A.}\
  \bibnamefont {Balseiro}}, \ and\ \bibinfo {author} {\bibfnamefont
  {G.}~\bibnamefont {Usaj}},\ }\href@noop {} {\bibfield  {journal} {\bibinfo
  {journal} {arXiv:1409.2482}\ } (\bibinfo {year} {unpublished})}\BibitemShut
  {NoStop}%
\bibitem [{\citenamefont {Diehl}\ \emph {et~al.}(2011)\citenamefont {Diehl},
  \citenamefont {Rico}, \citenamefont {Baranov},\ and\ \citenamefont
  {Zoller}}]{Diehl11}%
  \BibitemOpen
  \bibfield  {author} {\bibinfo {author} {\bibfnamefont {S.}~\bibnamefont
  {Diehl}}, \bibinfo {author} {\bibfnamefont {E.}~\bibnamefont {Rico}},
  \bibinfo {author} {\bibfnamefont {M.~A.}\ \bibnamefont {Baranov}}, \ and\
  \bibinfo {author} {\bibfnamefont {P.}~\bibnamefont {Zoller}},\ }\href@noop {}
  {\bibfield  {journal} {\bibinfo  {journal} {Nature Physics}\ }\textbf
  {\bibinfo {volume} {7}},\ \bibinfo {pages} {971} (\bibinfo {year}
  {2011})}\BibitemShut {NoStop}%
\bibitem [{\citenamefont {Budich}\ \emph {et~al.}(shed)\citenamefont {Budich},
  \citenamefont {Zoller},\ and\ \citenamefont {Diehl}}]{Budich14}%
  \BibitemOpen
  \bibfield  {author} {\bibinfo {author} {\bibfnamefont {J.~C.}\ \bibnamefont
  {Budich}}, \bibinfo {author} {\bibfnamefont {P.}~\bibnamefont {Zoller}}, \
  and\ \bibinfo {author} {\bibfnamefont {S.}~\bibnamefont {Diehl}},\
  }\href@noop {} {\bibfield  {journal} {\bibinfo  {journal} {arXiv:1409.6341}\
  } (\bibinfo {year} {unpublished})}\BibitemShut {NoStop}%
\bibitem [{\citenamefont {Uhlmann}(1986)}]{Uhlmann86}%
  \BibitemOpen
  \bibfield  {author} {\bibinfo {author} {\bibfnamefont {A.}~\bibnamefont
  {Uhlmann}},\ }\href@noop {} {\bibfield  {journal} {\bibinfo  {journal} {Rep.
  Math. Phys.}\ }\textbf {\bibinfo {volume} {24}},\ \bibinfo {pages} {229}
  (\bibinfo {year} {1986})}\BibitemShut {NoStop}%
\bibitem [{\citenamefont {Rivas}\ \emph {et~al.}(2013)\citenamefont {Rivas},
  \citenamefont {Viyuela},\ and\ \citenamefont {Martin-Delgado}}]{Delgado13}%
  \BibitemOpen
  \bibfield  {author} {\bibinfo {author} {\bibfnamefont {A.}~\bibnamefont
  {Rivas}}, \bibinfo {author} {\bibfnamefont {O.}~\bibnamefont {Viyuela}}, \
  and\ \bibinfo {author} {\bibfnamefont {M.~A.}\ \bibnamefont
  {Martin-Delgado}},\ }\href {\doibase 10.1103/PhysRevB.88.155141} {\bibfield
  {journal} {\bibinfo  {journal} {Phys. Rev. B}\ }\textbf {\bibinfo {volume}
  {88}},\ \bibinfo {pages} {155141} (\bibinfo {year} {2013})}\BibitemShut
  {NoStop}%
\bibitem [{\citenamefont {Viyuela}\ \emph {et~al.}(2014)\citenamefont
  {Viyuela}, \citenamefont {Rivas},\ and\ \citenamefont
  {Martin-Delgado}}]{Delgado14a}%
  \BibitemOpen
  \bibfield  {author} {\bibinfo {author} {\bibfnamefont {O.}~\bibnamefont
  {Viyuela}}, \bibinfo {author} {\bibfnamefont {A.}~\bibnamefont {Rivas}}, \
  and\ \bibinfo {author} {\bibfnamefont {M.~A.}\ \bibnamefont
  {Martin-Delgado}},\ }\href {\doibase 10.1103/PhysRevLett.112.130401}
  {\bibfield  {journal} {\bibinfo  {journal} {Phys. Rev. Lett.}\ }\textbf
  {\bibinfo {volume} {112}},\ \bibinfo {pages} {130401} (\bibinfo {year}
  {2014})}\BibitemShut {NoStop}%
\bibitem [{\citenamefont {Wang}\ \emph {et~al.}(2013)\citenamefont {Wang},
  \citenamefont {Steinberg}, \citenamefont {Jarillo-Herrero},\ and\
  \citenamefont {Gedik}}]{Gedik13}%
  \BibitemOpen
  \bibfield  {author} {\bibinfo {author} {\bibfnamefont {Y.~H.}\ \bibnamefont
  {Wang}}, \bibinfo {author} {\bibfnamefont {H.}~\bibnamefont {Steinberg}},
  \bibinfo {author} {\bibfnamefont {P.}~\bibnamefont {Jarillo-Herrero}}, \ and\
  \bibinfo {author} {\bibfnamefont {N.}~\bibnamefont {Gedik}},\ }\href
  {\doibase 10.1126/science.1239834} {\bibfield  {journal} {\bibinfo  {journal}
  {Science}\ }\textbf {\bibinfo {volume} {342}},\ \bibinfo {pages} {453}
  (\bibinfo {year} {2013})}\BibitemShut {NoStop}%
\bibitem [{\citenamefont {Onishi}\ \emph {et~al.}(shed)\citenamefont {Onishi},
  \citenamefont {Ren}, \citenamefont {Novak}, \citenamefont {Segawa},
  \citenamefont {Ando},\ and\ \citenamefont {Tanaka}}]{Onishi14}%
  \BibitemOpen
  \bibfield  {author} {\bibinfo {author} {\bibfnamefont {Y.}~\bibnamefont
  {Onishi}}, \bibinfo {author} {\bibfnamefont {Z.}~\bibnamefont {Ren}},
  \bibinfo {author} {\bibfnamefont {M.}~\bibnamefont {Novak}}, \bibinfo
  {author} {\bibfnamefont {K.}~\bibnamefont {Segawa}}, \bibinfo {author}
  {\bibfnamefont {Y.}~\bibnamefont {Ando}}, \ and\ \bibinfo {author}
  {\bibfnamefont {K.}~\bibnamefont {Tanaka}},\ }\href@noop {} {\bibfield
  {journal} {\bibinfo  {journal} {arXiv:1403.2492}\ } (\bibinfo {year}
  {unpublished})}\BibitemShut {NoStop}%
\bibitem [{\citenamefont {Karch}\ \emph {et~al.}(2010)\citenamefont {Karch},
  \citenamefont {Olbrich}, \citenamefont {Schmalzbauer}, \citenamefont {Zoth},
  \citenamefont {Brinsteiner}, \citenamefont {Fehrenbacher}, \citenamefont
  {Wurstbauer}, \citenamefont {Glazov}, \citenamefont {Tarasenko},
  \citenamefont {Ivchenko}, \citenamefont {Weiss}, \citenamefont {Eroms},
  \citenamefont {Yakimova}, \citenamefont {Lara-Avila}, \citenamefont
  {Kubatkin},\ and\ \citenamefont {Ganichev}}]{Karch10}%
  \BibitemOpen
  \bibfield  {author} {\bibinfo {author} {\bibfnamefont {J.}~\bibnamefont
  {Karch}}, \bibinfo {author} {\bibfnamefont {P.}~\bibnamefont {Olbrich}},
  \bibinfo {author} {\bibfnamefont {M.}~\bibnamefont {Schmalzbauer}}, \bibinfo
  {author} {\bibfnamefont {C.}~\bibnamefont {Zoth}}, \bibinfo {author}
  {\bibfnamefont {C.}~\bibnamefont {Brinsteiner}}, \bibinfo {author}
  {\bibfnamefont {M.}~\bibnamefont {Fehrenbacher}}, \bibinfo {author}
  {\bibfnamefont {U.}~\bibnamefont {Wurstbauer}}, \bibinfo {author}
  {\bibfnamefont {M.~M.}\ \bibnamefont {Glazov}}, \bibinfo {author}
  {\bibfnamefont {S.~A.}\ \bibnamefont {Tarasenko}}, \bibinfo {author}
  {\bibfnamefont {E.~L.}\ \bibnamefont {Ivchenko}}, \bibinfo {author}
  {\bibfnamefont {D.}~\bibnamefont {Weiss}}, \bibinfo {author} {\bibfnamefont
  {J.}~\bibnamefont {Eroms}}, \bibinfo {author} {\bibfnamefont
  {R.}~\bibnamefont {Yakimova}}, \bibinfo {author} {\bibfnamefont
  {S.}~\bibnamefont {Lara-Avila}}, \bibinfo {author} {\bibfnamefont
  {S.}~\bibnamefont {Kubatkin}}, \ and\ \bibinfo {author} {\bibfnamefont
  {S.~D.}\ \bibnamefont {Ganichev}},\ }\href {\doibase
  10.1103/PhysRevLett.105.227402} {\bibfield  {journal} {\bibinfo  {journal}
  {Phys. Rev. Lett.}\ }\textbf {\bibinfo {volume} {105}},\ \bibinfo {pages}
  {227402} (\bibinfo {year} {2010})}\BibitemShut {NoStop}%
\bibitem [{\citenamefont {Karch}\ \emph {et~al.}(2011)\citenamefont {Karch},
  \citenamefont {Drexler}, \citenamefont {Olbrich}, \citenamefont
  {Fehrenbacher}, \citenamefont {Hirmer}, \citenamefont {Glazov}, \citenamefont
  {Tarasenko}, \citenamefont {Ivchenko}, \citenamefont {Birkner}, \citenamefont
  {Eroms}, \citenamefont {Weiss}, \citenamefont {Yakimova}, \citenamefont
  {Lara-Avila}, \citenamefont {Kubatkin}, \citenamefont {Ostler}, \citenamefont
  {Seyller},\ and\ \citenamefont {Ganichev}}]{Karch11}%
  \BibitemOpen
  \bibfield  {author} {\bibinfo {author} {\bibfnamefont {J.}~\bibnamefont
  {Karch}}, \bibinfo {author} {\bibfnamefont {C.}~\bibnamefont {Drexler}},
  \bibinfo {author} {\bibfnamefont {P.}~\bibnamefont {Olbrich}}, \bibinfo
  {author} {\bibfnamefont {M.}~\bibnamefont {Fehrenbacher}}, \bibinfo {author}
  {\bibfnamefont {M.}~\bibnamefont {Hirmer}}, \bibinfo {author} {\bibfnamefont
  {M.~M.}\ \bibnamefont {Glazov}}, \bibinfo {author} {\bibfnamefont {S.~A.}\
  \bibnamefont {Tarasenko}}, \bibinfo {author} {\bibfnamefont {E.~L.}\
  \bibnamefont {Ivchenko}}, \bibinfo {author} {\bibfnamefont {B.}~\bibnamefont
  {Birkner}}, \bibinfo {author} {\bibfnamefont {J.}~\bibnamefont {Eroms}},
  \bibinfo {author} {\bibfnamefont {D.}~\bibnamefont {Weiss}}, \bibinfo
  {author} {\bibfnamefont {R.}~\bibnamefont {Yakimova}}, \bibinfo {author}
  {\bibfnamefont {S.}~\bibnamefont {Lara-Avila}}, \bibinfo {author}
  {\bibfnamefont {S.}~\bibnamefont {Kubatkin}}, \bibinfo {author}
  {\bibfnamefont {M.}~\bibnamefont {Ostler}}, \bibinfo {author} {\bibfnamefont
  {T.}~\bibnamefont {Seyller}}, \ and\ \bibinfo {author} {\bibfnamefont
  {S.~D.}\ \bibnamefont {Ganichev}},\ }\href {\doibase
  10.1103/PhysRevLett.107.276601} {\bibfield  {journal} {\bibinfo  {journal}
  {Phys. Rev. Lett.}\ }\textbf {\bibinfo {volume} {107}},\ \bibinfo {pages}
  {276601} (\bibinfo {year} {2011})}\BibitemShut {NoStop}%
\bibitem [{\citenamefont {Caldeira}\ and\ \citenamefont
  {Leggett}(1981)}]{Caldiera81}%
  \BibitemOpen
  \bibfield  {author} {\bibinfo {author} {\bibfnamefont {A.~O.}\ \bibnamefont
  {Caldeira}}\ and\ \bibinfo {author} {\bibfnamefont {A.~J.}\ \bibnamefont
  {Leggett}},\ }\href {\doibase 10.1103/PhysRevLett.46.211} {\bibfield
  {journal} {\bibinfo  {journal} {Phys. Rev. Lett.}\ }\textbf {\bibinfo
  {volume} {46}},\ \bibinfo {pages} {211} (\bibinfo {year} {1981})}\BibitemShut
  {NoStop}%
\bibitem [{\citenamefont {Fregoso}\ \emph {et~al.}(2013)\citenamefont
  {Fregoso}, \citenamefont {Wang}, \citenamefont {Gedik},\ and\ \citenamefont
  {Galitski}}]{Gedik13b}%
  \BibitemOpen
  \bibfield  {author} {\bibinfo {author} {\bibfnamefont {B.~M.}\ \bibnamefont
  {Fregoso}}, \bibinfo {author} {\bibfnamefont {Y.~H.}\ \bibnamefont {Wang}},
  \bibinfo {author} {\bibfnamefont {N.}~\bibnamefont {Gedik}}, \ and\ \bibinfo
  {author} {\bibfnamefont {V.}~\bibnamefont {Galitski}},\ }\href {\doibase
  10.1103/PhysRevB.88.155129} {\bibfield  {journal} {\bibinfo  {journal} {Phys.
  Rev. B}\ }\textbf {\bibinfo {volume} {88}},\ \bibinfo {pages} {155129}
  (\bibinfo {year} {2013})}\BibitemShut {NoStop}%
\bibitem [{\citenamefont {Kohler}\ \emph {et~al.}(2005)\citenamefont {Kohler},
  \citenamefont {Lehmann},\ and\ \citenamefont {H\"{a}nggi}}]{Hanggi2005}%
  \BibitemOpen
  \bibfield  {author} {\bibinfo {author} {\bibfnamefont {S.}~\bibnamefont
  {Kohler}}, \bibinfo {author} {\bibfnamefont {J.}~\bibnamefont {Lehmann}}, \
  and\ \bibinfo {author} {\bibfnamefont {P.}~\bibnamefont {H\"{a}nggi}},\
  }\href@noop {} {\bibfield  {journal} {\bibinfo  {journal} {Phys. Rep.}\
  }\textbf {\bibinfo {volume} {406}},\ \bibinfo {pages} {379} (\bibinfo {year}
  {2005})}\BibitemShut {NoStop}%
\bibitem [{\citenamefont {Langemeyer}\ and\ \citenamefont
  {Holthaus}(2014)}]{Holthaus14}%
  \BibitemOpen
  \bibfield  {author} {\bibinfo {author} {\bibfnamefont {M.}~\bibnamefont
  {Langemeyer}}\ and\ \bibinfo {author} {\bibfnamefont {M.}~\bibnamefont
  {Holthaus}},\ }\href {\doibase 10.1103/PhysRevE.89.012101} {\bibfield
  {journal} {\bibinfo  {journal} {Phys. Rev. E}\ }\textbf {\bibinfo {volume}
  {89}},\ \bibinfo {pages} {012101} (\bibinfo {year} {2014})}\BibitemShut
  {NoStop}%
\end{thebibliography}

\end{document}